%% file: main.tex
\documentclass{article}
\pdfoutput=1
\usepackage[utf8]{inputenc}
\usepackage{hyperref}
 \hypersetup{ 
     colorlinks=true, 
     linkcolor=blue, 
     filecolor=black, 
     citecolor = red,       
     urlcolor=black, 
     } 
\usepackage{graphicx}
\usepackage{amsfonts,amssymb,amsmath,amsthm,amstext}
\usepackage{ dsfont }
\usepackage[usenames,dvipsnames]{xcolor}
\usepackage{fullpage}
\usepackage{tikz}
\usepackage[subpreambles]{standalone} 
\usetikzlibrary{patterns,positioning,arrows,calc,decorations.pathreplacing,intersections}
\usepackage{enumitem}
\usepackage[numbers]{natbib}
\usepackage{xspace} 

\newcommand{\argmax}{\mathrm{argmax}}
\newcommand{\pdif}[2]{\frac{\partial #1}{\partial #2}}
\newcommand{\reals}{\mathbb{R}}
\newcommand{\E}{\mathbb{E}}
\newcommand{\profile}{\mathbf{s}}
\newcommand{\sminus}{\profile_{-i}}
\newcommand{\one}{\mathds{1}}
\newcommand{\vprofile}{\mathbf{v}}

\newcommand{\Omit}[1]{}

\newcommand{\dnote}[1]{{\color{purple}#1}}

\newtheorem{theorem}{Theorem}[section]
\newtheorem{claim}[theorem]{Claim}
\newtheorem{observation}[theorem]{Observation}
\newtheorem{definition}[theorem]{Definition}
\newtheorem{corollary}[theorem]{Corollary}
\newtheorem{example}[theorem]{Example}
\newtheorem{lemma}[theorem]{Lemma}
\newtheorem{proposition}[theorem]{Proposition}
\newcommand{\Linear}{Auto-linear\xspace} 
\newcommand{\linear}{auto-linear\xspace} 

\DeclareMathOperator{\Wel}{\textsf{Wel}}

\title{Interdependent Public Projects\thanks{{This work was partially funded by the European Research Council (ERC) under the European Union's Horizon 2020 research and innovation program (grant agreement No. 866132).}}}

\author{Avi Cohen\\Tel Aviv University\thanks{chn.avi@gmail.com} \and
Michal Feldman\\Tel Aviv University\thanks{mfeldman@tauex.tau.ac.il} \and
Divyarthi Mohan\\Tel Aviv University\thanks{divyarthim@tau.ac.il} \and
Inbal Talgam-Cohen\\Technion\thanks{italgam@cs.technion.ac.il}}

\begin{document}

\begin{titlepage}
\maketitle
\begin{abstract}

In the interdependent values (IDV) model introduced by Milgrom and Weber [1982], agents have private signals that capture their information about different social alternatives, and the valuation of every agent is a function of all agent signals. 
While interdependence has been mainly studied for auctions, it is extremely relevant for a large variety of social choice settings, including the canonical setting of public projects. The IDV model is very challenging relative to standard independent private values, and welfare guarantees have been achieved through two alternative conditions known as {\em single-crossing} and {\em submodularity over signals (SOS)}. 
In either case, the existing theory falls short of solving the public projects setting. 

Our contribution is twofold: (i)~We give a workable characterization of truthfulness for IDV public projects for the largest class of valuations for which such a characterization exists, and term this class \emph{decomposable valuations}; (ii)~We provide possibility and impossibility results for welfare approximation in public projects with SOS valuations.
Our main impossibility result is that, in contrast to auctions, no universally truthful mechanism performs better for public projects with SOS valuations than choosing a project at random. Our main positive result applies to {\em excludable} public projects with SOS, for which we establish a constant factor approximation similar to auctions. Our results suggest that exclusion may be a key tool for achieving welfare guarantees in the IDV model.
\end{abstract}
\end{titlepage}

\input{intro}
\input{prelims}
\input{multi-linear}
\input{approximation}

\input{general-charac}

\bibliographystyle{plainnat}
\bibliography{bibliography}
 
\newpage
\appendix
\input{appendix}

\end{document}

%% file: intro.tex
\section{Introduction}

\noindent{\bf The interdependent values (IDV) model.} 
An underlying assumption in the vast majority of studies on mechanism design and social choice is that agents have independent private values \emph{(IPV)} for the different outcomes. 
And yet, in real-life scenarios, this is rarely the case. Indeed, in many auction settings, agent values are highly interdependent. A typical example is an auction for modern artwork, where an agent's value depends on others' assessment of the work's merit -- both since others may be better-informed on modern art, and since their opinions may determine the resale value of the work. 
Another canonical example is an auction for drilling rights, in which the information one
agent has about whether or not there is oil to be found is extremely relevant
to how another agent evaluates the rights being auctioned. 

The work of \citet{MW82} builds upon~\citet{Wilson77} to introduce the {\em interdependent value} model \emph{(IDV)}, which captures such interdependencies of values among the agents. 
This work was recently recognized by the Nobel prize in economics, awarded in 2020 to Milgrom and Wilson~\citep{Nobel20}. 
In the IDV model, every agent~$i$ has a privately-known signal $s_i$ that captures the agent's information about the different outcomes, in addition to a publicly-known valuation function that maps the signals of \emph{all} $n$ agents to $i$'s values for the outcomes.
The importance of this model is in providing a much more accurate depiction of valuations in practice -- for example, it sheds light on well-known phenomena like the winner's curse, which cannot be explained under IPV~\cite[e.g.,][]{ChenEW22}. 

\vspace{1ex}
\noindent{\bf IDV and welfare maximization.} 
The IDV model also raises fascinating theoretical challenges:
An important theme of algorithmic game theory is the interplay between truthful implementability and the approximation factor achievable for a certain optimization problem. In IPV this has been extensively studied for the objective of minimizing makespan, with a recent breakthrough showing a large gap between non-truthful and truthful approximation (even with no computational limitations)~\citep{ChristodoulouKK21}. For arguably the most natural objective -- welfare maximization -- there is no gap in IPV due to the VCG mechanism (computational considerations aside), which applies beyond auction settings to general social choice. In IDV however, such a gap exists even for welfare in auction settings.

The economics literature has studied conditions under which welfare maximizing auction design is possible in the IDV model. 
For single-dimensional settings, if valuations satisfy a condition called {\em single-crossing}, one can obtain the maximum welfare in an (ex-post) truthful mechanism \citep{Ausubel99}.
Recently, work in computer science has combined approximation with alternative assumptions to extend positive welfare results beyond single-crossing. \citet{EFFGK19} achieve a constant-factor approximation for valuations satisfying {\em submodularity over signals (SOS)}. They extend their result under \emph{separability} even to multi-dimensional settings. 

\vspace{1ex}
\noindent{\bf Public projects with IDV.} 
Beyond auction design, for more general social choice settings, the state of affairs is less clear. A well-studied such setting is the \emph{combinatorial public projects problem (CPPP)}, in which a set of resources is chosen to \emph{collectively} serve a community~\citep{BuchfuhrerSS10,SchapiraS08,LucierSST13,Markakis17,Dughmi11,DughmiRY16}. 
This captures various real-world resources ranging from roads to communication networks. In fact, CPPP can be shown to essentially capture any social choice setting (computational considerations aside -- see Section~\ref{sec:approximation}).
The motivation to study such settings is more relevant than ever given the current trend towards allowing communities more influence over public policies and public decision-making~\cite[e.g.,][]{BrandlBP+21}.
The purpose of this paper is to initiate the study of mechanism design for public projects under IDV. 

We briefly describe the CPPP setting and provide a running example (Example~\ref{eg:running}): There are $n=2$ agents and a pool of $m=3$ potential public projects, including building a new bridge, opening a library, and building a train station. The goal is, for a given $k\le m$, to choose which $k$ projects to realize in order to maximize social welfare. Every agent $i$ has a privately-known signal $s_i\in\reals^+$ capturing their private information about the projects. The value $v_{ij}$ of agent $i$ for project $j$ is an increasing function of all signals $s_1,\dots,s_n$. In Example~\ref{eg:running}, the values are simple linear functions, but more generally they can be \linear, polynomial, etc. 
Observe that despite the fact that the signals are single-dimensional, the value space of every agent is multi-dimensional (and the related complexities apply).

\begin{example}[Running example]
\label{eg:running}
We define a public projects instance with 
$n=2$ agents and $m=3$ potential projects, with the following values:
\begin{itemize}
    \item Agent 1's values: $v_{11} = 3s_2$, $v_{12} = \frac{s_1}{2}+ s_2$, $v_{13}=2s_1$.
    \item Agent 2's values: $v_{21} = s_2$, $v_{22} = s_1+ \frac{s_2}{2}$, $v_{23}=0$.
\end{itemize}
(Figure \ref{fig:three-linear-projects-with-single-crossing} in Section~\ref{sec:linear-characterization} depicts the values of agent 1 for the three projects as a function of her signal $s_1$, when agent 2's signal is fixed to $s_2=1$.)
\end{example}

\noindent{\bf Overview of our contribution.} Inspired by the study of welfare maximization for auctions with IDV, for public projects we explore the two known conditions under which positive results are attainable for auctions: single-crossing valuations, and SOS valuations. In both cases, the existing theory falls short of solving the public projects setting. 
Our contribution is twofold: (i)~We give useful characterizations of truthful mechanisms for public projects in IDV settings (Section~\ref{sec:linear-characterization}, complemented by Section~\ref{sec:general-charac}); (ii)~We provide possibility and impossibility results for welfare approximation in such settings (Section~\ref{sec:approximation} -- this stand-alone section can be referred to directly by the interested reader).
Beyond the concrete results, our study reveals interesting connections between properties (e.g.~single-crossing and W-Mon) that were previously studied separately in the truthful implementability literature. 


\subsection{Characterization of Implementability in IDV}
In much of the previous IDV literature the focus has been on single-dimensional settings and welfare maximization. It is well-known that in this case, the valuations need to satisfy a condition called single-crossing to obtain optimal welfare truthfully~\citep[e.g.][]{Maskin92,Ausubel99,DM2000,JehieM01}.
We go beyond both single-dimensional settings and the social welfare objective.

\vspace{1ex}
\noindent{\bf Single-crossing-based characterization for decomposable valuations. }
In Section~\ref{sec:linear-characterization} we develop a useful characterization as follows. We first study \emph{\linear} valuations -- for each agent $i$, valuation $v_i$ is linear as a function of $s_i$. We define a necessary and sufficient condition for truthful welfare maximization beyond single-dimensional settings, termed \emph{strong} single-crossing. This condition is based on a comparison of slopes between an agent's valuation and the social welfare of different outcomes. 
We can extend \linear to a general valuation class we refer to as \emph{decomposable}. A valuation is decomposable if the influence of agent $i$'s signal $s_i$ on her value $v_i$ is independent of the social choice outcome. This class of valuations encompasses many standard examples of interdependence studied in the literature, such as the resale model~\citep{MW82} or the wallet game~\citep{Klemperer1998}, and generalizes {previously studied classes of valuations such as fully linear valuations (for which~\citet{JehieM01} provide a characterization)} and separable valuations~\citep{MookherjeeR92} (for which \citet{CE02} give a characterization). 
Finally, to go beyond welfare maximization to any social choice function $f$ over decomposable valuations, we generalize strong single-crossing to $f$-single-crossing. To summarize:

\vspace{0.1in}
\noindent {\bf Theorem:} (See Theorem \ref{thm:linear-characterization})
For any decomposable valuation profile $\vprofile$, a social choice function $f$ is (ex-post IC-IR) implementable if and only if $\vprofile$ satisfies $f$-single-crossing.

\vspace{0.1in}
\noindent\textbf{Beyond decomposable valuations.} In Section~\ref{sec:general-charac} we first complement our useful characterization from Section~\ref{sec:linear-characterization} by providing a characterization beyond decomposable valuations.
{Namely, we show that \emph{weak} $f$-single-crossing is a necessary and sufficient condition for implementability in IDV settings. Like other general characterizations (e.g.~for general IPV social choice), this condition is not easy to utilize.} 

\begin{figure}
    \centering
        \input{EC 2022/figures/high-level-diagram.tex}
    \caption{A scheme of the connections between implementability characterizations.\\
    For example, the scheme shows the equivalence between $f$-single-crossing and W-Mon for decomposable valuations (see the ``triangle'' of arrows formed on the bottom-left). It further shows that the $f$-single-crossing property characterizes implementability for decomposable valuations in the IDV model (see the middle ``line'' of arrows from left to right). See Sections~\ref{sec:linear-characterization} and \ref{sec:general-charac} for details. 
    \\{\textsuperscript{\textdagger}Single-crossing is defined for single-dimensional settings. \\\textsuperscript{*}Strong single-crossing and $f$-single-crossing are defined for decomposable valuations (see Observation~\ref{obv:decomposable-ssc-fsc}).} 
    }
    \label{fig:high-level-scheme}
\end{figure}
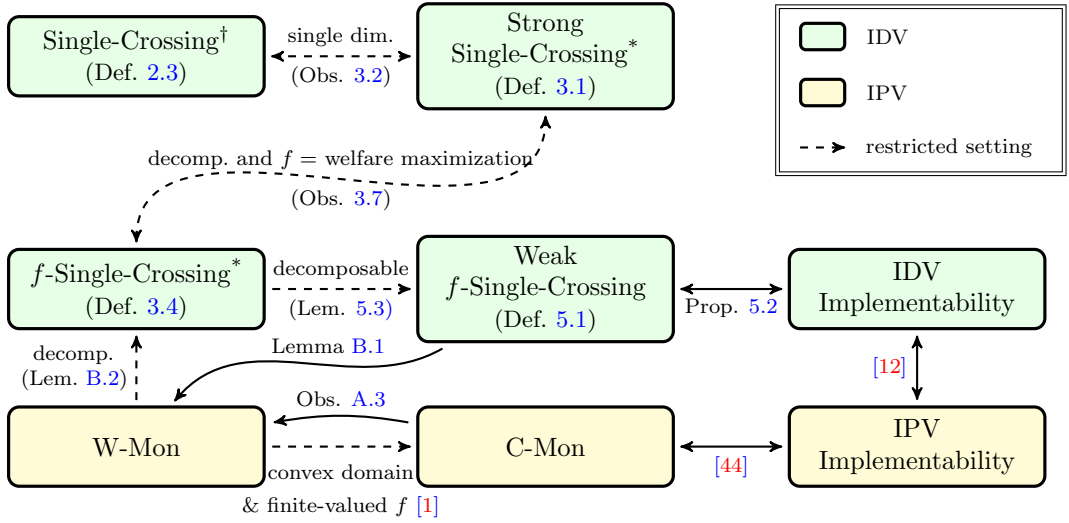

{We then study the connections between IPV and IDV implementability through the framework of~\citet{CE02}. Combining these with the results of Section~\ref{sec:linear-characterization} paints a comprehensive picture of the complex landscape of implementability.} 
The connections between the multitude of characterizations are summarized in Figure~\ref{fig:high-level-scheme}. Notably, our results establish that in the setting of decomposable valuations, \emph{all} the conditions depicted in Figure~\ref{fig:high-level-scheme} are equivalent. This provides an arguably surprising formal connection between single-crossing conditions for IDV, and W-Mon/C-Mon characterizations for IPV, which were previously studied separately in the literature. It also shows the importance of the class we have defined of decomposable valuations --- we prove the equivalence does not hold beyond it. In other words, decomposable valuations are the frontier for which a workable characterization of implementability exists, similar to convex domains for IPV which form the frontier between W-Mon (workable) and C-Mon characterizations.


\subsection{Approximate Welfare for Public Projects}
By our characterization results in Section~\ref{sec:linear-characterization}, for any public project instance with decomposable valuations, if strong single-crossing is satisfied then the optimal welfare can be obtained by a truthful mechanism. 
In Section~\ref{sec:approximation} we study settings without strong single-crossing.

Our first observation is that welfare maximization in auctions with $k$ identical items can be reduced to welfare maximization in public projects (choosing $k$ projects).
With this reduction in hand, known impossibility results in auction design (see \cite{EFFG18,EFFGK19}) immediately imply the following inapproximability results in public projects, even in cases where a single project should be chosen. 
First, no welfare approximation guarantee can be provided by any deterministic truthful mechanism.
Second, in the absence of additional constraints, no randomized mechanism can give any non-trivial (i.e., better than $1/m$) welfare approximation. 

\citet{EFFGK19} recently proposed to circumvent these impossibilities in auction design by considering valuations that satisfy a natural property called {\em submodularity over signals} (SOS).
Roughly speaking, these are valuations where the increase in one's value due to an increase in her signal is smaller when other signals are higher.
SOS is a natural condition that holds in essentially all examples in the IDV literature.
\citet{EFFGK19} showed that if valuations are  separable\footnote{{A valuation function $v_i : \mathcal{A}\times \mathcal{S} \to \reals^+$ is separable if there exist functions $h_i: \mathcal{A}\times S_i \to \reals^+ $ and $g_i: \mathcal{A}\times \mathcal{S}_{-i} \to \reals^+$ such that $v_i(a;\profile) = h_i(a;s_i) + g_i(a;\sminus)$}.} SOS, then one can obtain $4$-approximation even in general combinatorial auctions.

Does SOS come to our rescue in public projects as well?
{Unfortunately, SOS does not suffice for providing welfare guarantees in public project settings.}
Our main impossibility result is that even under separable SOS (linear) valuations, no  universally truthful mechanism can perform better than allocating a project at random. 

\vspace{0.1in}
\noindent {\bf Theorem:} (see Theorem~\ref{thm:lower-bound})
There exists a public projects instance with separable SOS valuations for which no universally truthful mechanism can give better than $1/m$ approximation to 
welfare.
\vspace{0.1in}

What is the source of the difference between auctions and public projects? 
A key component in the mechanism for auctions is a random partitioning of the agents into two groups of agents: those who are included and those who are excluded. 
The challenge in public projects is that once a project is in place, agents cannot be excluded from using it. 
However, an interesting subtype of public projects is the class of excludable public projects, often termed {\em club goods} in the economics literature \citep{Buchanan1965}. 
Examples of club goods include libraries, cinemas, swimming pools, or any public facility that benefits a restricted group of members. 

Our main positive result is that the truthful mechanism tailored for auctions in the IDV setting {\citep{EFFGK19}} can be adapted to excludable public projects.
This result greatly extends the mechanism's applicability.

\vspace{0.1in}
\noindent {\bf Theorem:} (see Theorem~\ref{thm:4-approx})
There is a universally truthful mechanism that gives $1/4$-approximation to welfare for excludable public projects with separable SOS valuations.
\vspace{0.1in}

A conceptual takeaway from our results is that exclusion is an important tool for welfare approximation guarantee in IDV settings.
In particular, our strong impossibility results may suggest that the strong notion of exclusion used in auction settings, where some agents are totally excluded, is inevitable.

\vspace{1ex}
\noindent{\bf Open problems.}
Our work suggests many natural directions for future research. 
\emph{First}, providing improved approximations or giving tighter bounds: Can we improve our $\frac{1}{4}$ approximation algorithm, perhaps by allowing randomized mechanisms that are truthful in expectation rather than universally truthful? Does the impossibility result of Theorem~\ref{thm:lower-bound} that necessitates excludability hold for truthful in expectation mechanisms? Can we show that the exclusion step at the beginning of the SOS mechanism is necessary not just for public projects but even for auctions?
\emph{Second}, relaxing assumptions: For example, is the separable SOS assumption necessary for a constant-factor approximation for public projects? (The same question is still open for auction design as well.)  
\emph{Third}, extending our results to related models: E.g., what is a characterization for truthfulness in IDV \emph{Bayesian} settings (i.e., with priors over the signals)? Can we design truthful mechanisms for other/additional objectives such as revenue maximization or budget balance? \emph{Fourth}, computational results: While our main focus is implementability rather than computation, we do get a polynomial-time truthful approximation algorithm for CPPP with IDV under additive valuations. A truthful solution to CPPP is APX-hard even for IPV for more general valuations~\citep{PSS08}. What approximation factor is tractable for CPPP with IDV beyond additive valuations?
%




\subsection{Additional Related Work}

\vspace{1ex}
\noindent{\bf Interdependence.}
\citet{CE02} show a connection between IPV and IDV truthfulness characterizations -- for completeness we include this as Lemma~\ref{lem:Chung-Ely} below. This enables importing to IDV characterizations for IPV, including the works of \cite{LMN03,BikhchandaniCS03,GuiMV04,SaksY05} on weak monotonicity (W-Mon). 

The bulk of algorithmic work on mechanism design has focused
on the IPV model. Early ventures beyond this model considered private values that are \emph{correlated} \citep[e.g.,][]{DobzinskiFK15}.
\citet{RT16} suggested to apply the computer science lens to the study of
mechanisms for interdependent values. They unified and generalized previous results to establish technical foundations for this study, including a characterization of truthful mechanisms for single-parameter settings with IDV. They also demonstrated natural sufficient conditions under which positive results in the form of robust mechanisms for approximate revenue maximization can be achieved. Concurrently, \citet{Li17} developed a simple near-optimal auction for revenue maximization with IDV, namely, the VCG mechanism with monopoly reserves, assuming monotone hazard rate value distributions.
\citet{ChawlaFK14} studied revenue approximation under relaxed assumptions.
They introduced a variant of the generalized VCG auction with reserve prices and random admission, and showed that this auction gives a constant approximation to the optimal expected revenue under a submodularity assumption on the valuation functions.

{There has also been work on multi-dimensional settings and multi-dimensional signals. \citet{JehieM01} study a general social choice setting with multi-dimensional signals 
under fully linear valuations where $v_{i}(a;\profile) = \sum_{j=1}^n \alpha_{ij}^a\cdot s_{ij}^a$ for each agent $i$. 
They consider a multi-dimensional signal space, compared to our single-dimensional signal space. 
On the other hand, they consider fully linear valuations, which is a special case of decomposable valuations considered in our work. They provide a characterization for Bayesian incentive compatibility, which translates to our $f$-single-crossing definition in the single-parameter signal regime (when applied to fully linear valuations). \citet{JehielMMZ06} showed that the constant social choice functions are the only deterministic social choice functions that are implementable in general multi-dimensional IDV settings with multi-dimensional signals and transferable utilities.}

The above studies and much of the other work on interdependent values assumed a single-crossing condition \citep[e.g.][]{MW82, dG82, Ausubel99, DM2000, BSV09, CKK15} (see also~\cite{WikiSingleCrossing22}).
Interdependent values with relaxed single-crossing were first studied by \citet{EFFG18} with a focus on welfare.
Interdependence without any single-crossing condition was studied by \citet{EFFGK19}, who introduced the submodularity over signals (SOS) condition. There is also a literature relaxing the knowledge assumption on the valuation functions themselves (in addition to the signals, which are privately-known by design in the IDV model). In \citep{DM2000,RobuPIJ13} the valuations are unknown only to the \emph{designer}; in \citep{EGZ22} they are unknown to the other agents as well (each agent knows only her own valuation). 

The described works on interdependence are in the context of auctions; we now turn to public projects which were studied for the IPV model.


\vspace{1ex}
\noindent{\bf Public projects.}
Public projects have long been studied in economics, with main objectives of welfare maximization and budget balancedness~\cite{Moulin94}.
\citet{PSS08} study the hardness of the Combinatorial Public Projects Problem (with independent, multi-parameter valuations), measured by both communication complexity and computational complexity.
A related but different body of AGT literature returns to the problem of cost sharing \citep{DMRS08, DO17} via approximation. In these settings each public project has a cost and the goal of the mechanism is to maximize welfare under the constraint that it must cover the costs of the chosen projects using the payments of the agents. 
The property of excludability is inherent to the model, since cost sharing implies that agents on which the mechanism does not impose a cost for a given project should not be able to enjoy the benefits of that project. 
The topic of excludable public goods was studied extensively in the economics literature \citep{DR99a, DR99b}, sometimes referred to as club goods \citep{Buchanan1965}. 
 Our paper extends this well-studied setting to multiple public goods and interdependent values (while we do not consider the cost of projects nor the aspect of budget-balancedness).

%% file: EC 2022/figures/high-level-diagram.tex
\begin{tikzpicture}[node distance = 2cm, auto,]
\colorlet{refcolor}{blue}

  \node[idv_block](single-crossing){Single-Crossing\textsuperscript{\textdagger}\\\small{(Def.~\textcolor{refcolor}{\ref{def:sc-auctions}})}};
  \node[idv_block, right=of single-crossing](strong-single-crossing) {Strong\\Single-Crossing\textsuperscript{*}\\\small{(Def.~\textcolor{refcolor}{\ref{defn:linear-single-crossing}})}}
    edge[<->, dashed] node[above]{single dim.} node[below]{(Obs. \textcolor{refcolor}{\ref{obv:single-sc and strong-sc}})}(single-crossing.east);
  \node[idv_block, below=of single-crossing](f-single-crossing){$f$-Single-Crossing\textsuperscript{*}\\\small{(Def.~\textcolor{refcolor}{\ref{defn:linear-f-single-crossing}})}}
    edge[<->, out=90, in=270, dashed] node[above]{decomp.~and $f=$ welfare maximization} node[below]{(Obs. \textcolor{refcolor}{\ref{obv:decomposable-ssc-fsc}})}(strong-single-crossing.south);
  \node[idv_block, right=of f-single-crossing](weak-f-single-crossing){Weak\\$f$-Single-Crossing\\\small{(Def.~\textcolor{refcolor}{\ref{defn:weak-single-crossing}})}}
    edge[<-,dashed]node[above]{decomposable}node[below]{(Lem. \textcolor{refcolor}{\ref{lem:sc-implies-weak-sc})} }(f-single-crossing.east);
  \node[idv_block, right=1.5cm of weak-f-single-crossing](idv-implementability){IDV\\Implementability}
    edge[<->]node[auto]{Prop. \color{refcolor}{\ref{prop:single-crossing-characterization}}}(weak-f-single-crossing.east);
  \node[ipv_block,below=1cm of f-single-crossing](wmon){W-Mon}
    edge[->,dashed, black, text=black, text width = 4em, text centered] node[auto] {\textcolor{black}{decomp.}\\ (Lem.~\textcolor{refcolor}{\ref{lem:wmon-implies-f-sc-when-decomp}})} (f-single-crossing.south)
    edge[<-, black, text=black, shorten >=4pt, shorten <= 2pt, in=210, out=50]node[above, xshift=1em]{Lemma \textcolor{refcolor}{\ref{lem:weak-f-sc-to-wmon}}}(weak-f-single-crossing);
  \node[ipv_block,right=of wmon](cmon){C-Mon}
    edge[<-, dashed] node[below=0.3em]{convex domain} node[below=1.5em]{\& finite-valued $f$ \textcolor{refcolor}{\citep{ABMH10}}} (wmon.east)
    edge[->, out = 170, in = 10, shorten >=3pt, shorten <= 3pt] node[above]{Obs. \textcolor{refcolor}{\ref{obv:cmon-implies-wmon}}} (wmon);
  \node[ipv_block,right=1.5cm of cmon](ipv-implementability){IPV\\Implementability}
    edge[<->] node[auto]{\textcolor{refcolor}{\citep{Rochet87}}}
    (cmon.east)
    edge[<->, black]node[auto, text=black]{\textcolor{refcolor}{\citep{CE02}}}(idv-implementability.south);
  \node[idv_block, rounded corners = 2pt, right=of strong-single-crossing.north, xshift = 1.37cm, yshift = -1.3em, text width = 1em, minimum height=1em](legend-idv){};
  \node[right=1ex of legend-idv, font=\footnotesize]{IDV};
  \node[ipv_block, rounded corners = 2pt, below=1em of legend-idv, text width = 1em, minimum height=1em](legend-ipv){};
  \node[right=1ex of legend-ipv, font=\footnotesize]{IPV};
  \node[rectangle, below=1em of legend-ipv, text width = 1em, minimum height=1em](dummy2){};
  \path (dummy2.west)  edge[->,dashed, shorten <=0pt, shorten >=0pt] (dummy2.east);
  \node[right=1ex of dummy2, font=\footnotesize] (legend-decomp){restricted setting};
  \path (dummy2.west |- legend-idv.north)+(-0.3,0.2) node(a1) {};
  \path (legend-decomp.south -| legend-decomp.east)+(0.3,-0.1) node(a2) {};
  \path[draw,double distance = 1pt] (a1) rectangle (a2);
\end{tikzpicture}

%% file: prelims.tex
\section{Preliminaries}
\label{sec:prelim}

In this section we introduce notation for classic social choice 
settings (Section~\ref{sub:prelim-IPV}) and define interdependent values (Section~\ref{sub:prelim-IDV}). 
We concisely summarize what's known in the literature about truthful implementation for interdependence (Section~\ref{sub:prelim-IC-IDV}; see Appendix~\ref{sub:prelim-IC-IPV} for truthful implementation \emph{without} interdependence). In Section~\ref{sub:prelim-PP} we present our main setting of interest -- public projects with interdependence.


\vspace{1ex}
\noindent{\bf Notation.}
For a vector $\mathbf{x}=(x_1,\ldots,x_n)$, we use the standard notation of $\mathbf{x}_{-i}$ to denote the same vector excluding $x_i$, and $(x'_i, \mathbf{x}_{-i})$ to denote the profile obtained by replacing $x_i$ with $x'_i$. 

\subsection{Independent Private Values (IPV)}
\label{sub:prelim-IPV}


\noindent{\bf Social choice settings.}
A social choice setting $(n,\mathcal{V},\mathcal{A})$ consists of $n$~\emph{agents} $\{1,\ldots,n\}$, a \emph{domain} of valuations $\mathcal{V}=V_1\times\ldots\times V_n$, and $\mu$ \emph{alternatives} $\mathcal{A}$ ($\mathcal{A}$ is also referred to as the \emph{outcome} space). In a particular instance, every agent~$i$ has a valuation function $v_i\in V_i$, 
where function $v_i : \mathcal{A} \to \reals^+$
specifies her value for every alternative (we sometimes also use a vector notation $v_i \in \reals^\mu$). 
A \emph{social choice function} $f:\mathcal{V}\to\mathcal{A}$ maps a valuation profile to one of the alternatives, possibly randomized.
We denote by $f_a(\vprofile)$ the probability assigned to alternative $a\in\mathcal{A}$.
A \emph{finitely-valued} $f$ has finitely-many distinct outcomes, that is, for every $i\in [n]$ and $\vprofile_{-i}$, $|\{ f(v_i,\vprofile_{-i}): v_i \in V_i\}| < \infty$.  
The \emph{welfare} of an alternative is the sum of the agents' values for it. 

\vspace{1ex}
\noindent{\bf Single-dimensional social choice settings.}
An important distinction is between settings with single- vs.~multi-dimensional domains. In the former,
the space $V_i$ is single-dimensional for every~$i$, i.e., there is a single real parameter that directly determines the valuation function $v_i$. 
Such domains are well-known to be significantly simpler for mechanism design than multi-dimensional ones.

Formally, a single-dimensional setting $(n,\mathcal{V},\mathcal{A})$ is a social choice setting in which the alternatives are subsets of agents, i.e., $\mathcal{A}\subseteq 2^{[n]}$. 
It is required that $\mathcal{A}$ be \emph{downward-closed} (if $W\in \mathcal{A}$ then for every $W'\subseteq W$, $W'\in \mathcal{A}$).
An outcome $W\in\mathcal{A}$ corresponds to a set of ``winning'' agents. For example, in a single-item auction, $\mathcal{A}=[n]$; in a multi-unit auction with $n$ units, $\mathcal{A}=2^{[n]}$. The valuations in such settings are simple:
slighly overloading notation, every agent $i$ has a single value $v_i$ for winning, such that
$v_i(W)=v_i$ if $i \in W$, and $v_i(W)=0$ otherwise. We use $f_i(\vprofile)$ to denote the probability that agent $i$ is a winner.

\vspace{1ex}
\noindent{\bf Mechanisms.}
A (direct revelation) mechanism solicits value reports $\mathbf{b} =(b_1,\ldots,b_n)$ from the agents. Its description is given by a pair $(f,p)$, where $f$ is a social choice function, and $p$ a collection of payment rules $p(\mathbf{b})=\{p_1(\mathbf{b}), \ldots, p_n(\mathbf{b})\}$.
Payment rule $p_i: \mathcal{V} \to \reals$ maps the bids $\mathbf{b}$ to the expected payment of agent $i$. For every $i$, agent $i$'s expected quasi-linear utility is given by 
$\sum_{a\in \mathcal{A}} f_a(\mathbf{b}) v_i(a) - p_i(\mathbf{b})$. We sometimes use an inner product  $\langle v_i, f(\mathbf{b})\rangle$ to denote $\sum_{a\in \mathcal{A}} f_a(\mathbf{b}) v_i(a)$.

\vspace{1ex}
\noindent{\bf Truthfulness and implementability.}
Truthfulness is without loss of generality by the revelation principle~\citep{Myerson1981}.
For standard IPV settings, we focus on the design of dominant-strategy incentive-compatible (IC) and individually rational (IR) mechanisms: 
A {deterministic} mechanism is considered \emph{truthful} if it is in every agent's best interest to participate and report her true value \emph{regardless of others' bids} (bidding truthfully is a dominant-strategy equilibrium of the mechanism).
Formally, a {deterministic} mechanism is dominant-strategy IC-IR if 
for every valuation profile $\vprofile\in \mathcal{V}$, agent $i\in[n]$ and bid $b_i\in V_i$,
\[
    \sum_{a\in\mathcal{A}} f_a(\vprofile) v_i(a) - p_i(\vprofile) \geq \max\Big\{ \sum_{a\in\mathcal{A}} f_a(b_i,\vprofile_{-i}) v_i(a) - p_i( b_i,\vprofile_{-i}), 0\Big\}. 
\]


For randomized mechanisms there are two levels of truthfulness: the mechanism can be truthful for every realization of its internal randomness, in which case we say it is \emph{universally truthful}; a weaker requirement is that the mechanism is  \emph{truthful in expectation} where the expectation is taken over its random coins. Unless stated otherwise, by ``truthful'' we refer to the former requirement of universal truthfulness. 

\begin{definition}
\label{def:implementable}
Consider a social choice setting $(n,\mathcal{V},\mathcal{A})$.
A social choice function $f$ is called \emph{implementable} if there exists a payment rule $p$ such that $(f,p)$ is truthful.
\end{definition}

\subsection{Interdependent Values (IDV)} 
\label{sub:prelim-IDV}

Our focus in this work is on the \emph{interdependent} values (IDV) model of \citet{MW82}. In the standard independent private values (IPV) model, the privately-known \emph{type} of each agent $i$ is her valuation~$v_i$. In IDV, however, the privately-known type is her \emph{signal} $s_i$, which captures her information on the social choice alternatives.%
\footnote{Note that this makes the types in our model -- despite its possible combinatorial nature -- single-parameter. We emphasize this does not mean that interdependent settings are always single-dimensional, rather that the dimensionality of the type is distinct from that of the setting.}
Formally, let the set of possible signals $S_i$ be $[0,1]$ for every agent $i\in[n]$. 
We denote by $\profile=(s_1,s_2,\ldots,s_n)$ a signal profile, and by $\mathcal{S} = \times_i S_i$ the signal space of the agents.
The \emph{values} of the agents are interdependent in the following sense: they depend not only on the chosen alternative $a\in\mathcal{A}$, but also on the information (signals) of all the agents. That is, for every $i$, the valuation function of agent $i$ is $v_i: \mathcal{A}\times \mathcal{S} \rightarrow \reals^+$ (we sometimes also denote $v_i(\cdot; \profile)$ as $v_i(\profile) \in \reals^\mu$ where the alternative $a$ is clear from the context). 
We assume (as is standard) that for every pair $i,i'\in[n]$, valuation $v_i$ is monotone non-decreasing in signal $s_{i'}$ . 
The collection of valuation functions $\vprofile=\{v_i\}_i$ is publicly known.

\vspace{1ex}
\noindent{\bf Social choice with IDV.}
Classic social choice settings can be adapted to IDV with only slight changes (mainly, $\profile$ replacing $\vprofile)$: A social choice setting with IDV $(n,\mathcal{S},\mathcal{A},\vprofile)$ consists of $n$ agents, a signal space~$\mathcal{S}$, alternatives $\mathcal{A}$, and a profile of $n$ valuation functions $v_i: \mathcal{A}\times \mathcal{S} \rightarrow \reals^+$. 
We refer to such settings as interdependent values (IDV) settings or as social choice settings with IDV.
A social choice function $f: \mathcal{S} \rightarrow \mathcal{A}$ maps each signal profile $\profile$ to a (possibly random) social alternative. For any distribution over outcomes $\delta \in \Delta(\mathcal{A})$,  we use $v_i(\delta; \profile)$, $\langle v_i(\profile) , \delta\rangle$, and $\sum_{a\in \mathcal{A}}\delta(a)v_i(a;\profile)$ interchangeably to denote agent $i$'s expected value.
A 
single-dimensional setting with IDV is similarly defined. For simplicity, in single-dimensional settings we often omit the alternative from the valuation function notation, and let $v_i:\mathcal{S} \rightarrow \reals^+$ be such that $v_i(a; \profile)=v_i(\profile)$ if $i \in a$ (agent $i$ wins), and $v_i(a; \profile) = 0$ otherwise.

\vspace{1ex}
\noindent{\bf Subclasses of interdependent valuations.} 
An important subclass of interdependent valuations are \emph{\linear}valuations. A valuation $v_i$ is \linear if for every $a$ and $\profile_{-i}$, $v_i(a; s_i, \profile_{-i})$ is linear as a function of $s_i$. This class encompasses well-studied special cases like the resale model~\citep{Myerson1981,RT16,EdenFTZ21} or Klemperer’s wallet game~\citep{Klemperer1998}.


We introduce the following 
natural extension, which distills the essence of \linear valuations and serves as the domain of our main results. We say that a valuation $v_i$ is decomposable if there exist functions $\hat v_i : \mathcal{S} \to \reals^+$, $h_i : \mathcal{A}\times \mathcal{S}_{-i} \to \reals^+$ and $g_i: \mathcal{A}\times \mathcal{S}_{-i} \to \reals$, such that for every $a,\,\profile$ we have 
\begin{equation}
v_i(a;\profile) = \hat{v}_i(\profile)\cdot h_i(a;\sminus) + g_i(a;\sminus).\label{eq:decomp}
\end{equation}

For example, the valuation function $v_1$ defined as $v_1(1;\profile) = s_1^2 + s_2$, $v_1(2;\profile) = s_1^2s_2$ is a decomposable valuation, where $\hat{v}_1(s_1,s_2) = s_1^2$, $h_1(1;s_2) = 1 ,h_1(2;s_2) = s_1$, $g_1(1;s_2) = s_2$, and $g_1(2;s_2)=0$.

We observe that decomposable valuations are strictly more general than the following classes of valuations: (i) separable\footnote{Not to be confused with separable SOS, as defined in~\citep{EFFGK19} -- see Section~\ref{sec:approximation}.} environments{~\citep{MookherjeeR92,CE02}}, where $h_i$ does not depend on $\sminus$ and $g_i \equiv 0$, (ii) single-dimensional settings,  where $h_i(a;\sminus) = 1$ if $i\in a$ and $0$ otherwise and $g_i\equiv 0$, and (iii) \linear valuations where $\hat v_i(s_i,\sminus)=s_i$.



\vspace{1ex}
\noindent{\bf Mechanisms and truthfulness with IDV.}
A mechanism for IDV solicits signal rather than value reports $\mathbf{b} =(b_1,\ldots,b_n)$ from the agents. 
For every $i$, agent $i$'s expected quasi-linear utility is given by 
$\sum_{a\in A} f_a(\mathbf{b}) v_i(a; \profile) - p_i(\mathbf{b})$.
In interdependent settings, it is well-known that we cannot hope to design mechanisms where truth-telling is a dominant strategy, therefore incentive compatibility and individual rationality are defined \emph{ex-post}. I.e., a mechanism for IDV is considered \emph{truthful} if it is in every bidder's best interest to participate and report her true signal \emph{given that all other agents bid truthfully} (i.e., bidding truthfully is an ex-post equilibrium of the mechanism).
Formally, a deterministic mechanism is ex-post IC-IR if for every signal profile $\profile\in\mathcal{S}$, agent $i\in[n]$ and bid $b_i\in S_i$,
\[ 
    \sum_{a\in \mathcal{A}} f_a(\profile) v_i(a; \profile) - p_i(\profile) \geq \max\Big\{ \sum_{a\in \mathcal{A}} f_a(b_i,\profile_{-i}) v_i(a;\profile) - p_i( b_i,\profile_{-i}), 0\Big\},
\]
where $f_a(\profile)$ indicates whether alternative $a$ is chosen given $\profile$.
As above, a randomized mechanism is universally truthful if it consists of a distribution over truthful deterministic mechanisms. 

\subsection{Implementability with IDV: What's Known}
\label{sub:prelim-IC-IDV}

In Appendix~\ref{sub:prelim-IC-IPV} we include several known key results on implementability for IPV. For IDV, implementability is more subtle, as we now detail.

\vspace{1ex}
\noindent{\bf Single-dimensional settings.}
For single-dimensional settings with IDV, monotonicity characterizes implementable social choice functions (similarly to IPV). Recall that 
$f_i(\profile)$ denotes the probability that agent~$i$ wins. Then:

\begin{theorem}[Implementability with IDV: Single-dimensional~(e.g.~\citep{RT16})]\label{thm:single-dim-charaterization}
For every single-dimensional IDV setting, a social choice function $f$ is (ex post IC-IR) implementable if and only if for every $i,\sminus$ 
it holds that $f_i(s_i,\sminus)$ is monotone non-decreasing in signal~$s_i$. 
\end{theorem}

Unlike IPV, it turns out that even to implement a social choice function like welfare maximization, with IDV an additional \emph{single-crossing} condition is needed in order to achieve monotonicity and hence truthfulness. 
Many definitions for such a condition appear in the literature, the following is adapted from~\citep{RT16}: 

\begin{definition}[Single-crossing condition]\label{def:sc-auctions}
Given a single-dimensional setting with IDV, we say that the valuation profile $\vprofile$ satisfies \emph{single-crossing} if for all agents $i,j$ and every  $ \profile \in \mathcal{S}$,
 $
    \pdif{v_i}{s_i}(\profile) \ge \pdif{v_j}{s_i}(\profile)
 $.
\end{definition}
\noindent That is, agent $i$'s signal influences her own valuation more than the valuation of any other agent. 
This definition or slight variations of it in effect characterize truthfulness,
as for every single-dimensional setting with IDV, 
welfare maximization is implementable
if and only if the valuation profile $\vprofile$ satisfies single-crossing.



\vspace{1ex}
\noindent{\bf Beyond single-dimensional settings.}
For settings beyond single-dimensional, there is a characterization for the following class in the context of welfare maximization:
When the valuation functions are separable,~\citet{CE02} define a strengthened single-crossing condition that is necessary and sufficient for welfare maximization. 
Similar single-crossing conditions have appeared in the welfare maximization context under varying assumptions~\citep[e.g.,][]{DM2000,ItoP06,RobuPIJ13}. 
We shall define a generalization of this condition in Definition~\ref{defn:linear-single-crossing} for the broader class of decomposable valuations, and apply it beyond welfare maximization.

To our knowledge, there is no generalized single-crossing condition that characterizes welfare maximization implementability in \emph{general} social choice settings with IDV.
However, \citet{CE02} establish a relationship between characterizing implementability in social choice settings with IDV and with IPV:

\begin{lemma}[Characterizing Implementability with IDV vs.~IPV, \citep{CE02}]
\label{lem:Chung-Ely}
Let $C$ be a condition that applies to a quadruple $(\mathcal{A}, S, v(\cdot), f)$ that represents a single-agent, single-dimensional social choice setting with alternative set $\mathcal{A}$, value domain $\{v(s)\mid s\in S\}$, and a social choice function $f$.
The following characterizations using condition $C$ are equivalent, i.e., one holds if and only if the other holds.
\begin{itemize}
    \item In a social choice setting with IDV, a social choice function $f : \mathcal{S} \rightarrow \mathcal{A}$ is ex post IC if (respectively only if) $\forall i, \forall s_{-i}$, the quadruple $(\mathcal{A}, S_i, v_i(\cdot~; s_{-i},\cdot), f( s_{-i},\cdot))$ satisfies condition $C$;
    \item 
    In a social choice setting with IPV, social choice function $f : \mathcal{V} \rightarrow \mathcal{A}$ is dominant-strategy IC if (respectively only if) $\forall i, \forall s_{-i}$, the quadruple $(\mathcal{A}, V_i, v_i, f(s_{-i},\cdot))$ satisfies condition $C$.
\end{itemize}
\end{lemma}

\subsection{Interdependent Public Projects}
\label{sub:prelim-PP}

A classic \emph{combinatorial public projects setting} $(n,\mathcal{V},m,k)$ is a multi-dimensional social choice setting in which there are $n$ agents, $m$ projects $\{1,\ldots,m\}$, and a number $k\le m$, $k \in \mathbb{N}$. 
The alternatives are all combinations of up to $k$ projects, i.e., $\mathcal{A}=\{T\subseteq[m]\mid |T| \le k\}$.
A combinatorial public projects setting \emph{with interdependent values} is described by $(n,\mathcal{S},\vprofile,m,k)$, and the value of agent $i$ for project set $T$ is $v_i(T,\profile)$. 
In the combinatorial public projects \emph{problem (CPPP)}, the input is a public projects setting and the objective is to find a subset $T^*$ of up to $k$ projects that maximizes the social welfare: $T^*\in \arg\max_{T\in\mathcal{A}}\sum_{i=1}^n v_i(T,\profile)$.

In the following example, the valuation of agent~1 is \linear while the valuation of agent~2 is not \linear (since $v_{22}$ depends on $(s_2)^2$) but is decomposable, and the valuation of agent~3 is not even decomposable (since $v_{31}$ and $v_{32}$ depend on $s_3$ in different ways -- recall Eq.~\eqref{eq:decomp}). 

\begin{example}[Public project instance with interdependence]
There are $n=3$ agents and $m=2$ projects. Assume $k=1$, i.e., a single project can be realized. 
Each signal space is $\{0, 1, 2\}$, and the valuations of the agents for the two projects depend on the signals as follows:
\begin{itemize}
    \item \textbf{Agent 1: \Linear.} $v_{11}=s_1s_2$, $v_{12}=s_1({s_2})^2$;
    \item \textbf{Agent 2: Decomposable.} $v_{21}=s_1 + (s_2)^2$, $v_{22}=s_1({s_2})^2$;
    \item \textbf{Agent 3: Non-decomposable.} $v_{31}=s_1 + s_2 + s_3$, $v_{32} = s_1(s_3)^3$.
\end{itemize}
Analysis: Under signal profile $(1,0,0)$, the welfare-maximizing project is project~1 (with social welfare of $2$, compared to a social welfare of $0$ for the other project). Under signal profile $(1,2,2)$, the welfare-maximizing project is project~2 (with social welfare of $16$, compared to a social welfare of $12$ for the other project). 
\end{example}

%% file: multi-linear.tex
\section{Implementability Characterization for Decomposable
Valuations} 
\label{sec:linear-characterization}
In this section we first study necessary and sufficient conditions for truthful implementation of a social choice function for IDV settings with \linear valuations. {In Section~\ref{sec:warm-up-welfare} we define a strong single-crossing property that is necessary and sufficient for truthful welfare maximization, and in Section~\ref{sub:linear-characterization} we generalize this to a necessary and sufficient condition for implementability of any social choice function $f$. {We observe that the ``decomposability'' property of \linear valuations is what drives this characterization.} Indeed, in Section~\ref{sec:decomposable} we show that the characterization of Section~\ref{sub:linear-characterization} extends to the more general class of {\em decomposable} valuations.}

\subsection{Warm-Up and Illustrations: Welfare Maximization for Auto-linear Valuations}\label{sec:warm-up-welfare}
In this section {we study truthful welfare maximization} and provide intuition for our generalization of single-crossing towards a characterization of truthfulness.
%
We define a condition similar to single-crossing that is necessary and sufficient for truthful welfare maximization beyond single-dimensional settings. Our single-crossing condition, {termed} \emph{strong single-crossing}, is based on the comparison of the slopes of an agent's valuation and the slopes of the social welfare for different outcomes. For instance, in Example~\ref{eg:running}, the slopes of agent $1$'s valuations with respect to $s_1$ are $0,\frac{1}{2}$, and $2$ for the projects $1,2$, and $3$ respectively, and the slopes of the social welfare with respect to $s_1$ are $0,\frac{3}{2}$, and $2$ for the projects $1,2$, and $3$ respectively. Notice that, when considering the projects in increasing order with respect to either the slopes of the valuation or the slopes of the welfare we obtain the same ordering, i.e., $1<2<3$. This consistency in the order of slopes is the essence of our single-crossing condition (when it holds for each agent $i$ with respect to signal $s_i$).

{Other studies have considered similar strong single-crossing conditions for truthful welfare maximization, for example,~\citet{DM2000} define a stronger condition for general multi-item auctions,{~\citet{JehieM01} define this for general social setting under fully linear valuations},~\citet{CE02} define this for \emph{separable environments} (which is non-comparable to multi-item auctions), and~\citet{ItoP06} apply it to auctions with single-minded bidders (which is a separable environment).}


\begin{definition}[Strong single-crossing]\label{defn:linear-single-crossing}
 Given any IDV social choice setting $(n,\mathcal{S}, \mathcal{A},\vprofile)$ with \linear valuations $\vprofile$. Let $\Wel(a;\profile) = \sum_{i} v_i(a; \profile)$ denote the welfare of allocation $a\in \mathcal{A}$ given signal profile $\profile$.
 We say that $\vprofile$ satisfies strong single-crossing, if for each $i\in [n]$ and $\sminus\in \mathcal{S}_{-i}$
 \[
     \pdif{\Wel}{s_i}(a_1;\profile)\le \pdif{\Wel}{s_i}(a_2;\profile) \le \ldots \le \pdif{\Wel}{s_i}(a_r;\profile)
 \]
 implies
 \[
    \pdif{v_i}{s_i}(a_1;\profile)\le \pdif{v_i}{s_i}(a_2;\profile) \le \ldots \le \pdif{v_i}{s_i}(a_r;\profile),
 \]
 where $a_1,\ldots, a_r$ denote all the outcomes such that $a_j \in \max_{a\in \mathcal{A}} \Wel(a;z,\sminus)$ for some $z \in S_i$.
\end{definition}

\begin{observation}\label{obv:single-sc and strong-sc}
Applying the strong single-crossing condition to single-item auctions would mean that $\pdif{\Wel}{s_i}(i,\profile) \ge \pdif{\Wel}{s_i}(j,\profile)$ for all $i,j$, because $\pdif{v_i}{s_i}(i;\profile) \ge 0 = \pdif{v_i}{s_i}(j;\profile)$. Since by definition $\Wel(i;\profile) = v_i (i,\profile)$, that is exactly the same as the single-crossing condition $\pdif{v_i}{s_i}(i,\profile) \ge \pdif{v_j}{s_i}(j,\profile)$ from Definition~\ref{def:sc-auctions}.
\end{observation}

{Observe that for any \linear valuations $\vprofile$, for all outcomes $a$ and signal profile $\sminus$, the slopes with respect to $s_i$, $\pdif{v_i}{s_i}(a;\profile)$ and $\pdif{\Wel}{s_i}(a;\profile)$, is a constant with respect to $s_i$. 
}

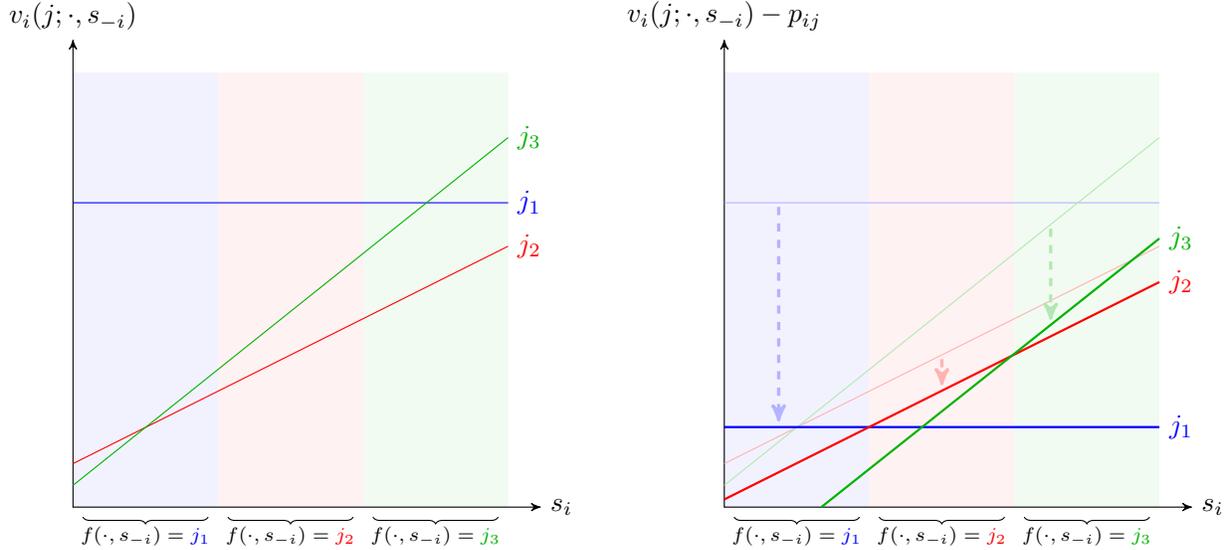
\begin{figure}
\begin{center}
    \resizebox{\textwidth}{!}{%
    \input{EC 2022/figures/three-projects.tex}
    }
\end{center}
    \caption{Truthful prices in an \linear setting with three projects where $f$-single-crossing holds.} 
    \label{fig:three-linear-projects-with-single-crossing}
\end{figure}

\vspace{1ex}
\noindent{\bf A visual illustration of strong single-crossing.}
The reason this condition implies implementability can be illustrated visually. 
Recall Example~\ref{eg:running}, where there are
$n=2$ agents and $m=3$ potential projects with the following values:
\begin{itemize}
    \item Agent 1's values: $v_{11} = 3s_2$, $v_{12} = \frac{s_1}{2}+ s_2$, $v_{13}=2s_1$.
    \item Agent 2's values: $v_{21} = s_2$, $v_{22} = s_1+ \frac{s_2}{2}$, $v_{23}=0$.
\end{itemize}
The plot in Figure \ref{fig:three-linear-projects-with-single-crossing} (left) depicts this for $i=1$ and $s_2=1$ ($j_\ell=\ell$ for $\ell\in[3]$), where $f$ is welfare maximization. The blue region ($s_1\in[0,5/3]$) is when project $1$ maximizes welfare; the red region ($s_1\in[5/3,3]$) is when project $2$ maximizes welfare; and the green region ($s_1 \geq 3$) is when project $3$ maximizes welfare.
Notice that if we order the projects by the slope of the welfare with respect to $s_1$ we get back the order of the projects that maximize welfare as $s_1$ increases. Moreover, the same order of slopes would be obtained when considering the valuation of agent $1$ for each of the projects. 
The fact that the slopes are aligned allows us to define prices such that the utility maximizing project is also the welfare maximizing project for each signal $s_1$. This is illustrated in the plot on the right, where the arrows depict the shift in the values induced by the prices.

\begin{figure}
\begin{center}
    \resizebox{\textwidth}{!}{%
    \input{EC 2022/figures/two-projects.tex}
    }%

\end{center}
    \caption{An \linear setting with two projects where $f$-single-crossing does not hold.} 
    \label{fig:two-linear-projects-without-single-crossing}
\end{figure}
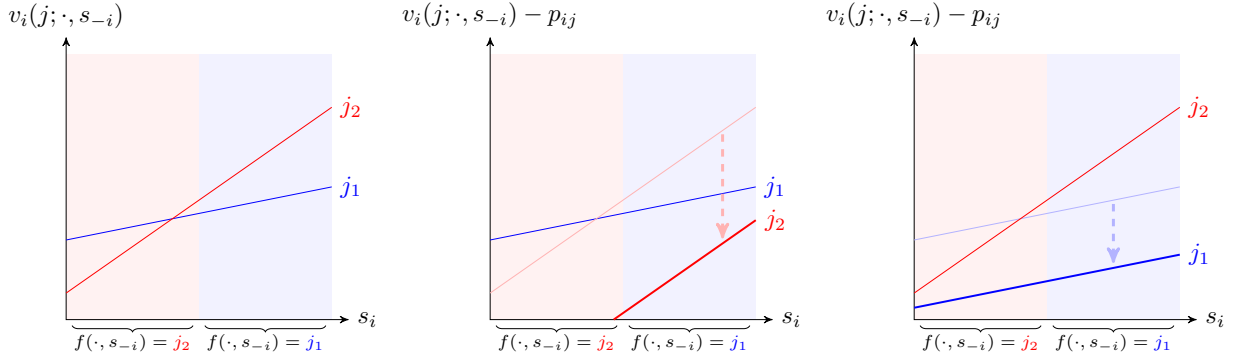

\vspace{1ex}
\noindent{\bf Strong single-crossing is necessary.}
Conversely, without having the slopes of $v_i$ with respect to $s_i$ ordered ``consistently'', i.e., when single-crossing does not hold, this is not possible. No set of prices can shift the lines such that the welfare maximizing outcome would also be the utility maximizer of agent $i$ for every signal $s_i$. This is illustrated in Figure \ref{fig:two-linear-projects-without-single-crossing}. The plot on the left depicts the utility of two projects as a function of $s_i$ (having fixed some set of signals $s_{-i}$ for the remaining agents). The blue and red lines depict $v_i$ for projects $j_1$ and $j_2$, respectively. The blue and red regions are the regions where project $j_1$ and $j_2$ maximizes welfare, respectively. Notice that the red line has a steeper slope than the blue line whereas the red region is to the left of the blue region --- the ordering is inconsistent! It is now easy to see that due to this discrepancy in the ordering, no set of prices can simultaneously place the blue line above the red line in the blue region and the red line above the blue line in the red region. Any price that shifts the red line below the blue line in the blue region would necessarily do the same in both regions. Similarly, a set of prices that ensure the red line is at the top in the red region would necessarily put it at the top in both regions. 

{Hence we get the following proposition characterizing efficient mechanisms for IDV settings with \linear valuations. We defer the proof to the appendix, where we provide a characterization for objective functions beyond welfare maximization.}

\begin{proposition}\label{prop:linear-characterization}
   For any IDV social choice setting $(n,\mathcal{S},\mathcal{A},\vprofile)$ with \linear valuations $\vprofile$, there is an ex post IC-IR mechanism that achieves optimal social welfare if and only if $\vprofile$ satisfies the strong single-crossing condition and the following payment identity holds for every $i$ and $\sminus$:
\begin{align*}
     p_{i}(x_j, \sminus) &= p_{i}(x_{j-1}, \sminus) + v_{i}(a_j; x_j) - v_{i}(a_{j-1};x_j) \\
     p_{i}(s_i, \sminus) &= p_{i}(x_j, \sminus) \quad \text{ for all } s_i \in [x_{j}, x_{j+1}) \\
     p_i(0,\sminus) &\le v_i(0,\sminus) 
\end{align*}
where $x_1 =0, x_2, \cdots, x_r \in S_i$ are the signals such that $f(\profile) = a_j$ for all $s_i \in [x_{j}, x_{j+1})$, and $a_1,\cdots, a_r$ as defined in Definition~\ref{defn:linear-single-crossing}.
\end{proposition}

\subsection{Beyond Welfare Maximization}
\label{sub:linear-characterization}
The simple demonstration above also illustrates the fact that the strong single-crossing condition need not be limited to welfare maximization exclusively. To find the correct prices, it is sufficient to consider the regions at which each project gets allocated.
The fact that the project was chosen by maximizing welfare had been incidental to our process. This allows us to define the \emph{$f$-single-crossing condition}, which characterizes implementablity for any social choice function $f$ over \linear valuations.\footnote{This also holds for decomposable valuations. See Section~\ref{sec:decomposable}.} {\citet{JehieM01} define similar conditions for characterizing Bayesian incentive compatibility for fully linear valuations.\footnote{{Their conditions apply to multi-parameter signal spaces, but they coincide with $f$-single-crossing in the special case of single-parameter settings with fully linear valuations.}}}

\begin{definition}[$f$-single-crossing]\label{defn:linear-f-single-crossing}
 Given any IDV social choice setting $(n,\mathcal{S}, \mathcal{A},\vprofile)$ with \linear valuations $\vprofile$, and a (possibly randomized) social choice function $f$.
 We say that $\vprofile$ satisfies $f$-single-crossing, if for each $i\in [n]$ and $\sminus\in \mathcal{S}_{-i}$, and any signals $s_i < s'_i$
 \[
     f(s_i,\sminus) = a_1 , f(s'_i,\sminus) = a_2
 \]
 implies
 \[
    \pdif{v_i}{s_i}(a_1;\profile)\le \pdif{v_i}{s_i}(a_2;\profile) .
 \]
\end{definition}

\begin{observation}\label{obv:f-sc and strong-sc}
Let $\vprofile$ be an \linear valuation profile, and $f$ the welfare maximizing social choice function.
Then, $\vprofile$ satisfies $f$-single-crossing if and only if $\vprofile$ satisfies strong single-crossing.
\end{observation}




{For any \linear valuation function $v_i$,   $\pdif{v_i}{s_i}{(a;\profile)}$ is a constant. 
In particular, there exists a function $g_i: \mathcal{A}\times \mathcal S_{-i} \to \reals^+$ such that,
\begin{align}
    v_i(a;\profile) = s_i\cdot \pdif{v_i}{s_i}(a;\profile) + g_i(a;\sminus) \qquad \forall \profile \in \mathcal{S}, \; a \in \Delta(\mathcal{A}),\label{eq:decomposable-linear}
\end{align}
where neither  $\pdif{v_i}{s_i}(a;\profile)$ nor $g_i(a;\sminus)$ depends on $s_i$. This decomposability property is key in proving the following proposition.
}

\begin{proposition}\label{prop:linear-f-necessary}
For any IDV social choice setting with \linear valuations $\vprofile$, if a social choice {function} $f$ is ex-post truthfully implementable, then $\vprofile$ satisfies $f$-single-crossing.
\end{proposition}
\begin{proof}
Recall that $f$ is a ex-post truthfully implementable if for all $i\in [n]$ there exists a price function $p_i: \mathcal{S} \to \reals$ such that
\[
    \langle v_i(\profile), f(\profile)\rangle - p_i(\profile) \geq
    \langle v_i(\profile), f(s_i',\sminus)\rangle - p_i(s_i',\sminus) \qquad \forall \profile \in \mathcal{S}, \; \forall s'_i \in S_i.
\]
Similarly, when the true signal is $s_i'$ we have,
\[
    \langle v_i(s_i',\sminus), f(s_i',\sminus)\rangle - p_i(s_i',\sminus) \geq
    \langle v_i(s_i',\sminus), f(\profile)\rangle - p_i(\profile) \qquad \forall \profile \in \mathcal{S}, \; \forall s'_i \in S_i.
\]
Adding the above two inequalities and rearranging we get,
\begin{align}
    \langle v_i(\profile) - v_i(s_i',\sminus), f(\profile)\rangle  \geq
    \langle v_i(\profile)- v_i(s_i',\sminus), f(s_i',\sminus)\rangle  \qquad \forall \profile \in \mathcal{S}, \; \forall s'_i \in S_i.
\end{align}
Since $\vprofile$ is \linear, we note that $v_i(\profile) - v_i(s'_i,\sminus) = (s_i - s'_i)\pdif{v_i}{s_i}(\profile)$ by using the decomposition of $v_i$ from Eq.~\eqref{eq:decomposable-linear}. This implies, for all $i\in [n]$, $\sminus \in \mathcal{S}_{-i}$, and $s_i > s'_i \in S_i$ ,
\begin{align*}
    \langle (s_i - s'_i)\pdif{v_i}{s_i}(\profile), f(\profile)\rangle  &\geq
    \langle (s_i - s'_i)\pdif{v_i}{s_i}(\profile), f(s_i',\sminus)\rangle
\end{align*}
and therefore
\begin{align*}
    \langle \pdif{v_i}{s_i}(\profile), f(\profile)\rangle  &\geq
    \langle \pdif{v_i}{s_i}(\profile), f(s_i',\sminus)\rangle  
\end{align*}
Hence proving $\vprofile$ satisfies $f$-single-crossing.
\end{proof}

\subsection{Extension to Decomposable Valuations}\label{sec:decomposable}


{Recall that the decomposition property 
of \linear valuations (Eq.~\ref{eq:decomposable-linear}) was key in enabling the characterization through $f$-single-crossing. 
With this in mind, we define a broader class of valuations called \emph{decomposable valuations} and  extend the  characterization found for \linear to this braoder class.

}


For any decomposable valuation $v_i$, there exists a decomposition of $v_i(a;\profile) = \hat{v_i}(\profile)h_i(a;\sminus) + g_i(a;\sminus)$, where $\hat v_i: \mathcal{S} \to \reals^+$, $h_i: \mathcal{A}\times \mathcal{S}_{-i} \to \reals^+$, and $g_i: \mathcal{A}\times \mathcal{S}_{-i} \to \reals^+$.
It is easy to see that the class of decomposable valuations also contains \linear valuations, where for any linear valuation $v_i$ the corresponding $\hat v_i (\profile) = s_i$.

For any decomposable valuations $\vprofile$, we note that $\pdif{v_i}{s_i}(a;\profile) = h_i(a;\sminus)\pdif{\hat v_i}{s_i}(\profile)$. Thus, Definitions~\ref{defn:linear-single-crossing} and~\ref{defn:linear-f-single-crossing} can be immediately extended to decomposable valuations.

\begin{observation}\label{obv:decomposable-ssc-fsc}
For any decomposable valuation $v_i$, we define \emph{strong single-crossing} and \emph{$f$-single-crossing} exactly like Definition~\ref{defn:linear-single-crossing} and Definition~\ref{defn:linear-f-single-crossing} respectively.
Further, for any decomposable valuation profile $\vprofile$ Observation~\ref{obv:f-sc and strong-sc} holds.
\end{observation}

Hence, by following the proof approach of Proposition~\ref{prop:linear-f-necessary} we immediately get that $f$-single-crossing is necessary {to truthfully implement $f$} for any IDV setting with decomposable valuations.

\begin{corollary}\label{cor:f-sc-necessary}
Given any IDV social choice setting with decomposable valuations $\vprofile$, a social choice function is truthfully implementable only if $\vprofile$ satisfies $f$-single-crossing. 
\end{corollary}

Moreover, we show that, for any decomposable valuation profile $\vprofile$, $f$-single-crossing is both necessary and sufficient condition for ex-post truthful implementability. {We defer the proof to the appendix.}


\begin{theorem}\label{thm:linear-characterization}
    For any IDV social choice setting with decomposable valuations $\vprofile$, a mechanism {$(f,p)$} is ex-post IC-IR  if and only if for every $i$, $\sminus$, $f$-single-crossing holds, and the following payment identity and payment inequality hold:
    \begin{align*}
        p_i(\profile) &= \; p_i(0,\sminus) + \int_{0}^{s_i} \left\langle v_i(t,\sminus), \pdif{f}{s_i}(t,\sminus) \right\rangle dt ;  \\
        p_i(0,\sminus) &\leq  \left\langle v_i(0,\sminus), f(0,\sminus) \right\rangle 
    \end{align*}
    
    Moreover, if we additionally require prices to be non-negative, it is sufficient to additionally have the following payment identity:
    $
        p_i(0,\sminus) =  \left\langle v_i(0,\sminus), f(0,\sminus) \right\rangle
    $
\end{theorem}

%% file: EC 2022/figures/three-projects.tex
   \begin{tikzpicture}[scale=1.2]
    \colorlet{color_one}{blue}
    \colorlet{color_two}{red}
    \colorlet{color_three}{black!30!green}

    \coordinate (y) at (0,5) ;
    \coordinate (x) at (5,0);
    \coordinate (top) at ($0.93*(y)$);
    \coordinate (region-one) at ($.31*(x)$);
    \coordinate (region-two) at ($.62*(x)$);
    \coordinate (region-three) at ($.93*(x)$);
    \draw[fill=color_one!5, draw=none] (top) rectangle (region-one);
    \draw[fill=color_two!5, draw=none] let \p1=(top), \p2=(region-one), \p3=(region-two) in
            (\x2,\y1) rectangle (\p3);
    \draw[fill=color_three!5, draw=none] let \p1=(top), \p2=(region-two), \p3=(region-three) in
            (\x2,\y1) rectangle (\p3);
    \draw[axis] (y) node[above]{$v_i(j;\cdot,s_{-i})$} -- (0,0) --  (x) node[right]{$s_i$};
    \draw[color_one] let \p1=(top), \p2=(region-three) in
        ($0.7*(\p1)$) -- ($(\x2,0.7*\y1)$) node[right]{$j_1$};
    \draw[color_two] let \p1=(top), \p2=(region-three) in
        ($0.1*(\p1)$) -- ($(\x2,0.6*\y1)$) node[right]{$j_2$};
    \draw[color_three] let \p1=(top), \p2=(region-three) in
        ($0.05*(\p1)$) -- ($(\x2,0.85*\y1)$) node[right]{$j_3$};
    \draw[decorate,decoration={brace, mirror}] let \p1=(0,0), \p2=(region-one) in
       ($(\x1+3pt,\y1-3pt)$) -- ($(\x2-3pt,\y2-3pt)$) node[below, midway, text centered, font=\scriptsize]{$f(\cdot,s_{-i})=\textcolor{color_one}{j_1}$};
    \draw[decorate,decoration={brace, mirror}] let \p1=(region-one), \p2=(region-two) in
        ($(\x1+3pt,\y1-3pt)$) -- ($(\x2-3pt,\y2-3pt)$) node[below, midway, text centered, font=\scriptsize]{$f(\cdot,s_{-i})=\textcolor{color_two}{j_2}$};
    \draw[decorate,decoration={brace, mirror}] let \p1=(region-two), \p2=(region-three) in
        ($(\x1+3pt,\y1-3pt)$) -- ($(\x2-3pt,\y2-3pt)$) node[below, midway, text centered, font=\scriptsize]{$f(\cdot,s_{-i})=\textcolor{color_three}{j_3}$};
  \end{tikzpicture}
\quad
  \begin{tikzpicture}[scale=1.2]
    \colorlet{color_one}{blue}
    \colorlet{color_two}{red}
    \colorlet{color_three}{black!30!green}

    \coordinate (y) at (0,5) ;
    \coordinate (x) at (5,0);
    \coordinate (top) at ($0.93*(y)$);
    \coordinate (region-one) at ($.31*(x)$);
    \coordinate (region-two) at ($.62*(x)$);
    \coordinate (region-three) at ($.93*(x)$);
    \draw[fill=color_one!5, draw=none] (top) rectangle (region-one);
    \draw[fill=color_two!5, draw=none] let \p1=(top), \p2=(region-one), \p3=(region-two) in
            (\x2,\y1) rectangle (\p3);
    \draw[fill=color_three!5, draw=none] let \p1=(top), \p2=(region-two), \p3=(region-three) in
            (\x2,\y1) rectangle (\p3);
    \draw[axis, name path=axis] (y) node[above]{$v_i(j;\cdot,s_{-i})-p_{ij}$} -- (0,0) --  (x) node[right]{$s_i$};
    \draw[color_one!30] let \p1=(top), \p2=(region-three) in
        ($0.7*(\p1)$) --node[very near start](arrow_one_start){} ($(\x2,0.7*\y1)$) node[right]{};
    \draw[color_two!30] let \p1=(top), \p2=(region-three) in
        ($0.1*(\p1)$) --node[midway](arrow_two_start){} ($(\x2,0.6*\y1)$) node[right]{};
    \draw[color_three!30] let \p1=(top), \p2=(region-three) in
        ($0.05*(\p1)$) --node[near end](arrow_three_start){} ($(\x2,0.85*\y1)$) node[right]{};
    \draw[->, dashed, very thick, color_one!30, shorten <= -2pt, shorten >=2pt] (arrow_one_start) -- +(0,-2.4);
    \draw[->, dashed, very thick, color_two!30, shorten <= -2pt, shorten >=2pt] (arrow_two_start) -- +(0,-.383);
    \draw[->, dashed, very thick, color_three!30, shorten <= -2pt, shorten >=2pt] (arrow_three_start) -- +(0,-1.08);
    \draw[color_one, thick] let \p1=(top), \p2=(region-three) in
        ($0.7*(\p1)-(0,2.4)$) -- ($(\x2,0.7*\y1)-(0,2.4)$) node[right]{$j_1$};
    \draw[color_two, thick] let \p1=(top), \p2=(region-three) in
        ($0.1*(\p1)-(0,.383)$) -- ($(\x2,0.6*\y1)-(0,.383)$) node[right]{$j_2$};
    \draw[color_three, thick] let \p1=(top), \p2=(region-three) in
        (1.0375,0) -- ($(\x2,0.85*\y1)-(0,1.08)$) node[right]{$j_3$};
    
    \draw[decorate,decoration={brace, mirror}] let \p1=(0,0), \p2=(region-one) in
       ($(\x1+3pt,\y1-3pt)$) -- ($(\x2-3pt,\y2-3pt)$) node[below, midway, text centered, font=\scriptsize]{$f(\cdot,s_{-i})=\textcolor{color_one}{j_1}$};
    \draw[decorate,decoration={brace, mirror}] let \p1=(region-one), \p2=(region-two) in
        ($(\x1+3pt,\y1-3pt)$) -- ($(\x2-3pt,\y2-3pt)$) node[below, midway, text centered, font=\scriptsize]{$f(\cdot,s_{-i})=\textcolor{color_two}{j_2}$};
    \draw[decorate,decoration={brace, mirror}] let \p1=(region-two), \p2=(region-three) in
        ($(\x1+3pt,\y1-3pt)$) -- ($(\x2-3pt,\y2-3pt)$) node[below, midway, text centered, font=\scriptsize]{$f(\cdot,s_{-i})=\textcolor{color_three}{j_3}$};
  \end{tikzpicture}

%% file: EC 2022/figures/two-projects.tex
   \begin{tikzpicture}[scale=0.8]
    \colorlet{color_one}{blue}
    \colorlet{color_two}{red}

    \coordinate (y) at (0,5) ;
    \coordinate (x) at (5,0);
    \coordinate (top) at ($0.94*(y)$);
    \coordinate (region-one) at ($.47*(x)$);
    \coordinate (region-two) at ($.94*(x)$);
    \draw[fill=color_two!5, draw=none] (top) rectangle (region-one);
    \draw[fill=color_one!5, draw=none] let \p1=(top), \p2=(region-one), \p3=(region-two) in
            (\x2,\y1) rectangle (\p3);
    \draw[axis] (y) node[above]{$v_i(j;\cdot,s_{-i})$} -- (0,0) --  (x) node[right]{$s_i$};
    \draw[color_one] let \p1=(top), \p2=(region-two) in
        ($0.3*(\p1)$) -- ($(\x2,0.5*\y1)$) node[right]{$j_1$};
    \draw[color_two] let \p1=(top), \p2=(region-two) in
        ($0.1*(\p1)$) -- ($(\x2,0.8*\y1)$) node[right]{$j_2$};
    \draw[decorate,decoration={brace, mirror}] let \p1=(0,0), \p2=(region-one) in
       ($(\x1+3pt,\y1-3pt)$) -- ($(\x2-3pt,\y2-3pt)$) node[below, midway, text centered, font=\scriptsize]{$f(\cdot,s_{-i})=\textcolor{color_two}{j_2}$};
    \draw[decorate,decoration={brace, mirror}] let \p1=(region-one), \p2=(region-two) in
        ($(\x1+3pt,\y1-3pt)$) -- ($(\x2-3pt,\y2-3pt)$) node[below, midway, text centered, font=\scriptsize]{$f(\cdot,s_{-i})=\textcolor{color_one}{j_1}$};
  \end{tikzpicture}
  \begin{tikzpicture}[scale=0.8]
    \colorlet{color_one}{blue}
    \colorlet{color_two}{red}

    \coordinate (y) at (0,5) ;
    \coordinate (x) at (5,0);
    \coordinate (top) at ($0.94*(y)$);
    \coordinate (region-one) at ($.47*(x)$);
    \coordinate (region-two) at ($.94*(x)$);
    \draw[fill=color_two!5, draw=none] (top) rectangle (region-one);
    \draw[fill=color_one!5, draw=none] let \p1=(top), \p2=(region-one), \p3=(region-two) in
            (\x2,\y1) rectangle (\p3);
    \draw[axis] (y) node[above]{$v_i(j;\cdot,s_{-i})-p_{ij}$} -- (0,0) --  (x) node[right]{$s_i$};
    \draw[color_one] let \p1=(top), \p2=(region-two) in
        ($0.3*(\p1)$) --  ($(\x2,0.5*\y1)$) node[right]{$j_1$};
    \draw[color_two!30] let \p1=(top), \p2=(region-two) in
        ($0.1*(\p1)$) -- node[very near end](arrow_two_start){} ($(\x2,0.8*\y1)$) node[right]{};
    \draw[->, dashed, very thick, color_two!30, shorten <= -2pt, shorten >=2pt] (arrow_two_start) -- +(0,-2);

    \draw[color_two, thick] let \p1=(top), \p2=(region-two) in
    (2.1857162,0) --  ($(\x2,0.8*\y1)-(0,2)$) node[right]{$j_2$};
    \draw[decorate,decoration={brace, mirror}] let \p1=(0,0), \p2=(region-one) in
    ($(\x1+3pt,\y1-3pt)$) -- ($(\x2-3pt,\y2-3pt)$) node[below, midway, text centered, font=\scriptsize]{$f(\cdot,s_{-i})=\textcolor{color_two}{j_2}$};
    \draw[decorate,decoration={brace, mirror}] let \p1=(region-one), \p2=(region-two) in
     ($(\x1+3pt,\y1-3pt)$) -- ($(\x2-3pt,\y2-3pt)$) node[below, midway, text centered, font=\scriptsize]{$f(\cdot,s_{-i})=\textcolor{color_one}{j_1}$};
  \end{tikzpicture}
  \begin{tikzpicture}[scale=0.8]
    \colorlet{color_one}{blue}
    \colorlet{color_two}{red}

    \coordinate (y) at (0,5) ;
    \coordinate (x) at (5,0);
    \coordinate (top) at ($0.94*(y)$);
    \coordinate (region-one) at ($.47*(x)$);
    \coordinate (region-two) at ($.94*(x)$);
    \draw[fill=color_two!5, draw=none] (top) rectangle (region-one);
    \draw[fill=color_one!5, draw=none] let \p1=(top), \p2=(region-one), \p3=(region-two) in
            (\x2,\y1) rectangle (\p3);
    \draw[axis] (y) node[above]{$v_i(j;\cdot,s_{-i})-p_{ij}$} -- (0,0) --  (x) node[right]{$s_i$};
    \draw[color_one!30] let \p1=(top), \p2=(region-two) in
        ($0.3*(\p1)$) -- node[near end](arrow_one_start){} ($(\x2,0.5*\y1)$) node[right]{};
    \draw[color_two] let \p1=(top), \p2=(region-two) in
        ($0.1*(\p1)$) -- ($(\x2,0.8*\y1)$) node[right]{$j_2$};
    \draw[->, dashed, very thick, color_one!30, shorten <= -2pt, shorten >=2pt] (arrow_one_start) -- +(0,-1.2);

    \draw[color_one, thick] let \p1=(top), \p2=(region-two) in
        ($0.3*(\p1)-(0,1.2)$) -- ($(\x2,0.5*\y1)-(0,1.2)$) node[right]{$j_1$};
    \draw[decorate,decoration={brace, mirror}] let \p1=(0,0), \p2=(region-one) in
    ($(\x1+3pt,\y1-3pt)$) -- ($(\x2-3pt,\y2-3pt)$) node[below, midway, text centered, font=\scriptsize]{$f(\cdot,s_{-i})=\textcolor{color_two}{j_2}$};
    \draw[decorate,decoration={brace, mirror}] let \p1=(region-one), \p2=(region-two) in
     ($(\x1+3pt,\y1-3pt)$) -- ($(\x2-3pt,\y2-3pt)$) node[below, midway, text centered, font=\scriptsize]{$f(\cdot,s_{-i})=\textcolor{color_one}{j_1}$};
  \end{tikzpicture}

%% file: approximation.tex
\section{Approximate Welfare Maximization in Public Projects}
\label{sec:approximation}
 
 

In this section we focus on the public projects setting with IDV valuations, and study truthful mechanisms that get approximately-optimal social welfare. 
Recall that we are given $n$ agents and $m$ projects, and we want to allocate up to $k$ projects. We observe in Section~\ref{sub:reduction} that even with $k= 1$, the public projects setting is rich enough to capture any general social choice setting (computation aside). This is because, given any social choice setting with IDV valuations $(n,\mathcal{S},\mathcal{A},\vprofile)$, we can consider a public project setting where each alternative $a\in \mathcal{A}$ has a corresponding project (that is, we have $n$ agents, $m = |\mathcal{A}|$ projects, and wish to choose a single project/alternative). 

By Proposition~\ref{prop:linear-characterization} it holds that for decomposable valuations, strong single-crossing is both a necessary and sufficient condition to achieve optimal social welfare truthfully. {We shall show in Section~\ref{sec:impossibility-result} that, in the absence of strong single-crossing, and even when restricting our attention to decomposable valuations (or even \linear valuations), there are strong impossibility results for obtaining non-trivial welfare guarantees truthfully.}

In Section~\ref{sub:exclusion-to-rescue} we apply a known tool for truthfulness -- excluding agents. Intuitively, excluded agents have no stake in the outcome and thus would report truthfully. 
Shifting focus to a variant of public goods that allows agent exclusion enables us to circumvent the impossibility result of Theorem~\ref{thm:lower-bound}.


\paragraph{Section preliminaries.}
We revisit well-studied conditions on valuation functions in the interdependent setting. In the IDV auctions setting, these have been useful for  obtaining approximately-optimal welfare truthfully even when single-crossing 
{does not hold}. The definitions below are adapted from~\cite{EFFGK19}.

\begin{definition}[Submodular-over-signals valuations]
A valuation function $v: \mathcal{A}\times\mathcal{S} \rightarrow \reals^+$ satisfies {\em submodularity over signals} (SOS) if for all $a\in\mathcal{A},\;\profile,\,\profile'\in \mathcal{S}$ and $i \in [n]$, where $\profile'$ is coordinate wise larger or equal to $\profile$ it holds that
\[
    v(a;s'_i,\sminus) - v(a;\profile) \ge v(a;\profile') - v(a; s_i,\profile'_{-i})
\]
\end{definition}
 
\begin{definition}[Separable SOS valuations]
For each agent $i$, a valuation function $v_i: \mathcal{A}\times\mathcal{S} \rightarrow \reals^+$ is  {\em separable SOS} if there exist functions $g_{-i}: \mathcal{A}\times\mathcal{S}_{-i} \rightarrow \reals^+$ and $h_i: \mathcal{A}\times S_i \rightarrow \reals^+$ such that 
\[
    v_i(a;\profile) = h_i(a;s_i) + g_{-i}(a; \profile_{-i}) ,
\]
 where $h_i(\cdot)$ and $g_{-i}(\cdot)$ are both weakly increasing and $g_{-i}(\cdot)$ is an SOS function.
\end{definition}



\subsection{Reducing Auctions to Public Projects and Impossibility Implications}
\label{sub:reduction}

In this section we show that known impossibility results from IDV single-item auctions carry over to IDV public projects via a simple reduction.


{
A single-item auction instance is given by $(n, \mathcal{S},\mathcal{A},\vprofile)$, where a single item needs to be allocated to one of $n$ agents, $\mathcal{A} = [n]$ is the set of possible outcomes (outcome $i$ denotes an allocation to agent $i$), every agent $i$ has a valuation $v_i: \mathcal{S} \rightarrow \reals^+$, where the value of agent $i$ for winning the item under signal profile  $\profile$ is $v_i(\profile)$, and $0$ otherwise (i.e., if $i$ is not the winner).
}
%

Given a single-item auction instance $(n, \mathcal{S},\mathcal{A},\vprofile)$, we define a public projects instance $(n,\mathcal{S},\vprofile',n, 1)$ with $n$ agents,  $m=n$ projects and $k=1$ (i.e., a single project should be chosen), where the valuation function $v'_i: [n]\times\mathcal{S} \rightarrow \reals^+$ is defined as
\[
    v'_i(j; \profile)= \begin{cases}
                          v_i(\profile) &\text{ if } j = i,\\
                          0 &\text{ otherwise }
                        \end{cases}
\]
That is, in the reduced instance, for every $j=1,\ldots,n$, we have a project $j$ associated with the outcome ``$j$ wins the item'', and only agent $j$ has non-zero value for this project. 
Thus, the social welfare of allocating a project $j$ (possibly, at random) in the public projects setting equals the social welfare of allocating the item to bidder $j$ in the corresponding auction setting, and the reduction is approximation preserving. 
As a corollary, the impossibility results for auction settings from~\citep{EFFG18} carry over to  public project settings. 




\vspace{1ex}
\noindent{\bf Known lower bounds for single-item auctions.} 
The following example shows that in the absence of single-crossing, no deterministic mechanism can obtain any approximation guarantee for the optimal social welfare, even under SOS valuations.


\begin{example}[No bound for deterministic mechanisms \citep{EFFG18}]
    \label{eg:impossiblity-deterministic}
    Consider a single-item auction with two agents. 
    Only agent $1$ has a signal, denoted by $s_1\in \{0,1\}$. The valuations of the agents for winning the item are
    \begin{align*}
        v_1(s_1) &= 1 + s_1, &
        v_2(s_1) &=H\cdot s_1,
    \end{align*}
    where $H$ is arbitrarily large. If the item is not allocated to agent $1$ when $s_1 = 0$, then the approximation ratio is infinite. On the other hand, if the item is allocated to agent $1$ when $s_1 = 0$, then by weak monotonicity, agent $1$ must also be allocated at $s_1=1$, leading to a $2/H$ fraction of the maximal social welfare.
\end{example}

One can easily verify that in Example~\ref{eg:impossiblity-deterministic}, no randomized mechanism can give better than $1/2$-approximation.

Beyond SOS valuations, no non-trivial approximation can be obtained by any truthful mechanism, as the next example shows.
\begin{example}[Lower bound of $n$ for randomized mechanisms without SOS \citep{EFFG18}]
    \label{eg:impossiblity-without-exclusion}
    Consider a single-item auction with $n$ agents. 
    For every agent $i$, $s_i\in \{0,1\}$, and the valuation of agent $i$
 for winning the item is
 \[
        v_i( \profile)=\prod_{j\neq i} s_j + \varepsilon\cdot s_i. 
    \]
    That is, agent $i$'s value for the item is non-negligible if and only if the signals $\sminus$ are $(1,\dots,1)$. When all signals are $1$, let $i$ be the agent who is allocated the item with probability at most $1/n$ (there must be such an agent in any feasible outcome). By weak monotonicity, at $\profile'=(\vec 1_{-i}, 0_i)$ the probability that agent $i$ is allocated the item is also at most $1/n$. Therefore, the welfare obtained is at most $\frac{1}{n}+\epsilon\cdot(n-1)$ at signal profile $\profile'$, while the optimal welfare is $1$, thus giving a factor-$n$ gap when $\varepsilon \to 0$.
\end{example}

\subsection{SOS to the Rescue?}
\label{sec:impossibility-result}

   
      In this section we show that public projects with interdependent values impose a unique challenge that does not arise in auction settings. 
   In combinatorial auctions with interdependent values, the SOS property (combined with separability) comes to our rescue; in particular, there exists a universally truthful mechanism that gives $1/4$-approximation to the optimal welfare for any instance with separable SOS valuations~\citep{EFFGK19}. 
   In stark contrast, the following theorem shows that in public projects, no {universally} truthful mechanism guarantees more than a $1/m$-approximation, even for separable SOS valuations.

\begin{theorem}\label{thm:lower-bound}
There exist linear valuation functions for which no ex-post IC-IR mechanism can perform better than allocating a project at random, i.e., we cannot get better than a $1/m$-approximation to the optimal social welfare.
\end{theorem}

\begin{proof}

Consider the following public projects instance, inspired by Example~\ref{eg:impossiblity-deterministic}.
   \begin{example}\label{eg:lower-bound-1/n}
    Consider $n$ agents, $m = n+1$ projects labelled $\{0 , 1, 2,\ldots, n\}$, and each agent $i\in [n]$ has a private signal $s_i \in \{0,1\}$ (using the convention that $s_{n+1} = s_1$). For any agent $i\in [n]$ and signals $\profile$, let $v_{i}(j; \profile)$ denote the value of agent $i$ for project $j$ under signals $\profile$. We define
    \[
    v_i (j; \profile) =
        \begin{cases}
        \varepsilon s_i + 1 & \text{ for }j =0,\\
        \frac{\varepsilon}{i+1} s_i + H^i\cdot s_{i+1} & \text{ for } j = i, \\
        \frac{\varepsilon}{j+1} s_i & \text{ otherwise}
        \end{cases}
    \]
    where $H$ is an arbitrarily large number.
   \end{example}
Recall that for a mechanism $(f,p)$ to be an $\alpha$-approximation (in the worst case) the (expected) welfare at each $\profile\in \{0,1\}^n$ needs to be at least an $\alpha$ factor of the optimal social welfare at $\profile$.
For $\profile = (0,0,\ldots, 0)$, project $0$ is optimal and the only outcome with positive welfare, hence any $\alpha$-approximation should allocate project $0$ with probability at least $\alpha$. For  $\profile = (1,1,\ldots,1)$ allocating project $n$ is optimal, and the welfare for any project other than $n$ is at most a $1/H$ factor of the optimal. For $\ell\le n-1$, $\profile = (0,1^{\ell},0^{n-\ell-1})$ allocating project $\ell$ is optimal, and the welfare for any project other than $\ell$ is at most a $1/H$ factor of the optimal. Hence when $\profile = (0,1^\ell,0^{n-\ell-1})$ (resp. $\profile = (1,1,\ldots,1)$), any $\alpha$-approximation should allocate project $\ell$ (resp. $n$) with probability at least $\alpha$. 

However, by Theorem~\ref{thm:linear-characterization}, for any truthfully implementable $f$ it holds that  $\vprofile$ satisfies $f$-single-crossing, since the valuations are linear (and hence decomposable).
Observe that, for all $i$ and $\sminus$ the slope $\pdif{v_i}{s_i}(j; \profile)$ is \emph{decreasing} in $j$. That is, for all $j_1 < j_2$ we have $v_{i} (j_1; s_i',\sminus) - v_{i} (j_1; \profile)> v_{i} (j_2; s_i',\sminus) - v_{i} (j_2; \profile)$, for $s_i' > s_i$. Thus, by $f$-single-crossing any truthfully implementable social choice function $f$ cannot both allocate $j_1$ at $\profile$ and $j_2 $ at $(s_i',\sminus)$ (it can only do one or the other). 

This implies, if $f$ allocates $\ell$ at $(0,1^\ell,0^{n-\ell-1})$ for $0\le \ell< n$, then at $(0,1^{\ell'},0^{n-\ell'-1})$ (resp. $(1,\ldots,1)$) it cannot allocate project $\ell'\neq\ell$ (resp. $n$). Recall, however, that for every project $0\le \ell< n$, project $\ell$ must be allocated with probability at least $\alpha$ at the signal $(0,1^\ell,0^{n-\ell-1})$, and by the previous claim these events must be disjoint. Therefore, with probability at least $n \cdot \alpha$ the social choice function $f$ allocates some project other than $n$ at the signal $(1,\ldots,1)$. Since the probability that project $n$ is allocated at $(1,\ldots,1)$ must be at least $\alpha$, this implies that $1-n\cdot\alpha \geq \alpha$. Therefore, $\alpha \leq 1/(n+1)$, and hence proving the required lower bound.

The last example shows that, in the absence of strong single-crossing, we cannot do any better than picking a project at random. Unlike auction settings, SOS does not come to our rescue.
\end{proof}

\subsection{Exclusion Allows Approximation} 
\label{sub:exclusion-to-rescue}

   


A key property that enables a universally truthful mechanism with constant-factor approximation in the auctions setting is {\em excludability}. 
Indeed, the mechanism in \citep{EFFGK19} entirely excludes a chosen set of agents from being allocated. 
This is a known tool for truthfulness~\cite{GoldbergHW01,EFFGK19}: the excluded agents have nothing to lose, and thus would report their valuations (or signals, in IDV settings) truthfully. 
Then, it only remains to show that by excluding agents cleverly enough, the loss in welfare is limited. 
Unfortunately, the situation with public projects is different.
In contrast to auctions, where goods are allocated to individual agents, pure public goods are non-excludable by definition.
Apparently, this challenge leads to the impossibility result cast in Theorem~\ref{thm:lower-bound}.
To alleviate this barrier, we turn to a variant of public goods that allows for exclusion. 




\vspace{1ex}
\noindent{\bf Excludable public projects.} 
We consider the setting of excludable public projects~\citep{DR99a, DR99b} (also known as ``club goods"). In this setting, for each project $j$ chosen by the mechanism, the mechanism is allowed to exclude a set of agents $E_j \subseteq [n]$ from using it. 
That is, agents in $E_j$ obtain no value from project $j$. 
%
In fact, our results hold even with respect to a more restricted class of mechanisms where $E_j = E_{j'}$ for every $j,j'\in[m]$; i.e., the set of excluded agents is identical for all projects.

Formally, an instance of \emph{globally excludable public projects} is defined by the tuple $(n, \mathcal{S},m,\vprofile, k)$, where a feasible outcome is given by a set $J \subseteq [m]$ of size at most $k$, along with a set $E\subseteq[n]$ of excluded agents. Slightly overloading notation, the valuation $v_i$ is defined as $v_i(J,E;\profile) = v_i(J;\profile)$  for $i\notin E$ and zero otherwise.


Global exclusion is useful for the design of truthful mechanisms. Indeed, as in the case of auctions, the excluded agents gain no benefit no matter what they report, and would therefore lose nothing from reporting their signals truthfully.

Notably, this is not the case for ``local exclusion", namely, where every project has a different set of excluded agents.
Indeed, an agent who is excluded from some project $j$ may still wish to misreport her signal in order to affect the allocation of a different project she is not excluded from.
Interestingly, even if agents have independent signals for different projects (this setting is beyond the one we consider here), an agent who is excluded from some project $j$ may still have incentive to lie about her signal for project $j$, since it may affect the allocation of different projects (see Appendix \ref{app:local-exclusion} for an example). 

Clearly, exclusion might harm welfare, as the excluded agents contribute nothing to the social welfare. 
Our main positive result is that the universally-truthful mechanism devised in \citep{EFFGK19} can be adapted to excludable public goods, and the same $1/4$-approximation to the optimal social welfare applies.

We first observe that it is without loss of generality (computational considerations aside) to restrict attention to the case where a single project should be chosen (i.e., $k=1$). 
Indeed, given an instance $(n,\mathcal{S},m,\vprofile,k)$ where $k$ projects should be chosen, one can consider an instance $(n,\mathcal{S},{m \choose k},\vprofile',1)$, where every $k$-project set in the original instance is a ``meta-project" in the new instance, and $v'_i(J,E;\profile) = v_i(J,E;\profile)$.


\vspace{1ex}
\noindent{\bf Approximation.}
We consider the \textsf{Random-Sampling-VCG} mechanism, introduced in~\cite{EFFGK19} for combinatorial auctions, adapted to excludable public projects as described below. 

For any setting with separable valuations, \textsf{Random-Sampling-VCG} is a universally truthful mechanism. This property arises from the fact that globally excluded agents have no incentive to lie about their signals as they get nothing in any possible outcome. Further, if the valuations are SOS, then using the key lemma (Lemma \ref{lem:key-lemma}), we get a $1/4$-approximation to the optimal welfare.

 \begin{theorem}\label{thm:4-approx}
 There is an ex-post IC-IR mechanism that guarantees a $1/4$-approximation for the optimal social welfare under the setting of excludable public projects with separable SOS valuations.
 \end{theorem}
 
We focus on globally excludable settings, as global exclusion is all that is required for this mechanism. Moreover, due to the reduction shown above, we may restrict our attention to settings with $k=1$.
We start by considering the following class of mechanisms, similar to the VCG-inspired mechanisms from ~\cite{EFFGK19}.

\begin{definition}[\textsf{$A$-exclusion-VCG}]
 Given any valuations $\vprofile$, and a subset of agents $A$, we define the \textsf{$A$-exclusion-VCG} mechanism as follows:
\begin{enumerate}
    \item All agents report some signals $\tilde{s_i}\in S_i$.
    \item For agents $i\notin A$, define $w_i(a;\tilde{s_i},\tilde{\mathbf{s}}_{A}) = v_i(a;\tilde{s}_i,\tilde{\mathbf{s}}_{A},\vec 0_{A^c\setminus \{i\}})$ for all projects $a$.
    \item Allocate $a^*\in \argmax_a \sum_{i\notin A} w_i(a;\tilde{s}_i,\tilde{\mathbf{s}}_{A})$.
    \item Compute generalized VCG prices for all $i\notin A$, 
        \[
            p_i(\tilde{\mathbf{s}}) = \left(g_{-i}(a^*;\tilde{\mathbf{s}}_{-i}) - g_{-i}(a^*;\tilde{\mathbf{s}}_{A},\vec 0_{A^c\setminus\{i\}})\right) - \sum_{i'\notin A\cup\{i\}} w_{i'}(a^*;\tilde{s}_{i'},\tilde{\mathbf{s}}_{A}) + \max_{a} \sum_{i' \notin A\cup\{i\}}w_{i'}(a;\tilde{s}_{i'},\tilde{\mathbf{s}}_{A})
        \]
\end{enumerate}
\end{definition}

\begin{lemma}\label{lem:A-exclusion-truthful}
The \textsf{$A$-exclusion-VCG} mechanism is truthful when the valuations are separable SOS\footnote{We actually do not need the full SOS property, we only need $v_i(a;\profile) = h_i(a;s_i) + g_{-i}(a;\sminus)$.}. 
\end{lemma}

\begin{proof}[Proof. {[Following \cite{EFFGK19}, Theorem 5.1]}]
%
For any agent $i$ with true signal $s_i$, and fixed $\tilde{s}_{-i} = \sminus$, we show that her utility for reporting $\tilde{s}_i = s_i$ is at least her utility for reporting $\tilde{s}_i = s'_i$ for all $s'_i\neq s_i$.

Clearly, for any $i\in A$, the utility of $i$ is $0$ (with no allocation and $0$ payment) for any reported signal. Hence truthfulness trivially follows.
Let $f(\profile)$ be the output of the $A$-exclusion mechanism for any reported signals $\tilde{\mathbf{s}} = \profile$. For any agent $i\notin A$, observe that the price $p_i(\tilde{\mathbf{s}})$ does not depend on $\tilde{s_i}$. For separable SOS valuations, with $v_i(a,\profile)= h_i(a;s_i) + g_{-i}(a;\sminus)$ we have $w_{i}(a;\tilde{s}_i,\tilde{\mathbf{s}}_A) = h_i(a; \tilde{s}_i) + g_{-i}(a;\tilde{\mathbf{s}}_A,\vec 0_{A^c\setminus \{i\}})$. Thus we see that for all $\tilde{s}_i$ and $\tilde{\mathbf{s}}_{-i} = \sminus$ and $\hat a=f(\tilde{s}_i,\sminus)$

\begin{align*}
    p_i(\tilde{s}_i,\sminus) &= \left(g_{-i}(\hat a;\sminus) - g_{-i}(\hat a;{\profile}_{A},\vec 0_{A^c\setminus\{i\}})\right) - \sum_{i'\notin A\cup\{i\}} w_{i'}(\hat a;{s}_{i'},{\profile}_{A}) + \max_{a} \sum_{i' \notin A\cup\{i\}}w_{i'}(a;{s}_{i'},{\profile}_{A}) \\
    &= \left(g_{-i}(\hat a;\sminus) - g_{-i}(\hat a;{\profile}_{A},\vec 0_{A^c\setminus\{i\}})\right) + w_{i}(\hat a;{s}_{i},{\profile}_{A}) - \sum_{i'\notin A} w_{i'}(\hat a;{s}_{i'},{\profile}_{A}) + \max_{a} \sum_{i' \notin A\cup\{i\}}w_{i'}(a;{s}_{i'},{\profile}_{A}) \\
    &= v_i(\hat a;\profile) - \sum_{i'\notin A} w_{i'}(\hat a;{s}_{i'},{\profile}_{A}) + \max_{a} \sum_{i' \notin A\cup\{i\}}w_{i'}(a;{s}_{i'},{\profile}_{A})
\end{align*}


Hence we have that the utility of agent $i$ with true signal $s_i$ and reported signal $\tilde{s}_i$ is, 
\[
  v_i(f(\tilde{s}_i,\sminus);\profile) - p_i(\tilde{s}_i,\sminus) =   \sum_{i'\notin A} w_{i'}(\hat a;{s}_{i'},{\profile}_{A}) - \left( \max_{a} \sum_{i' \notin A\cup\{i\}}w_{i'}(a;{s}_{i'},{\profile}_{A})\right)
\]

Note that, the first term is maximized at $\hat a = f(\profile)$ by definition of our social choice function. Thus, the utility of agent $i$ for $\tilde{s}_i = s_i$ is maximal.

Moreover, we note that the utility at $\tilde{s}_i = s_i$ is non-negative. This is because the first term is larger than the second term due to reason similar to VCG.
%
\end{proof}

The \textsf{$A$-exclusion-VCG} mechanism assumes an arbitrary set $A$ of excluded agents. This is used as a subroutine in the mechanism for which we obtain our approximation result, defined as follows.

\begin{definition}[\textsf{Random-Sampling-VCG}]\label{def:random-sampling-vcg}
Randomly sample a subset of agents $A\subseteq [n]$ uniformly at random. Then run the $A$-exclusion-VCG mechanism.
\end{definition}

By Lemma~\ref{lem:A-exclusion-truthful}, for every fixed $A\subseteq [n]$ the \textsf{$A$-exclusion-VCG} mechanism is truthful, therefore we obtain the following corollary.

\begin{corollary}~\label{cor:univ-truthful}
\textsf{Random-Sampling-VCG} is a universally truthful mechanism when the valuations are separable SOS.
\end{corollary}


The final component we need for the proof of the theorem is the Key Lemma (\citet{EFFGK19}).

\begin{lemma}[Key Lemma, ~\cite{EFFGK19}]\label{lem:key-lemma}
Let $v_i : {(\reals^+)}^n \rightarrow \reals^+$ be any SOS function. Let $A$ be a uniformly random subset of $[n]\setminus \{i\}$ and $B = A^c\setminus\{i\}$. Then we have for all $\profile$, 
$\E_A[v_i(s_i,\profile_A,\vec 0_B)] \ge \frac{1}{2}v_i(\profile).
$
\end{lemma}

We are now ready to prove the theorem. 

\begin{proof}[Proof of theorem~\ref{thm:4-approx}]
For every profile $\profile$, $i\in [n]$ and $a\in \mathcal{A}$, we have
 \[
     \E_A[w_i(a;s_i, \profile_A)\cdot\one_{\{i\notin A\}}] = \E[v_i(a;s_i,{\profile}_A,\vec 0_{A^c\setminus\{i\}})\mid i \notin A]\Pr[i \notin A]  \ge \frac{1}{4}v_i(a;\profile).
 \]
 
This follows from Lemma~\ref{lem:key-lemma}, because $v_i(a;\cdot)$ is an SOS function. Therefore, for any $\profile$ let $\tilde{a}$ be a welfare maximizing allocation. For every subset $A$, the social welfare of  \textsf{$A$-exclusion-VCG} is $\sum_{i\notin A} v_i(a^*;\profile) \ge \sum_{i\notin A} w_i(a^*;s_i,\profile_A)$ when $a^* \in \argmax_{a\in\mathcal{A}}\sum_{i\notin A} w_i(a;s_i,\profile_A)$.

Hence the social welfare of the \textsf{Random-Sampling-VCG} is at least


$$
    \E_{A}\Big[\max_{a\in \mathcal{A}}\sum_{i\notin A} w_i(a;s_i,\profile_A) \Big] \ge \E_{A}\Big[\sum_{i} w_i(\tilde{a};s_i,\profile_A)\cdot \one_{\{i\notin A\}}\Big] 
    = \sum_{i}\E_{A}[ w_i(\tilde{a};s_i,\profile_A)\cdot \one_{\{i\notin A\}}] 
     \ge \frac{1}{4} \sum_{i}v_i(\tilde{a};\profile).
$$
\end{proof}

%% file: general-charac.tex
\section{General Characterization of Social Choice Implementability}\label{sec:general-charac}

In this section we complement our characterization results from Section~\ref{sec:linear-characterization}. In Section~\ref{sub:beyond-decompose} we define \emph{weak} $f$-single-crossing and show that it characterizes implementability for general IDV settings. In Section~\ref{sec:idv-ipv-equiv} we discuss the connections between implementability in IPV and IDV, showing in particular that weak $f$-single-crossing is analogous to C-Mon and $f$-single-crossing is analogous to W-Mon. In Section~\ref{sub:impossibility-for-non-decomposable} we identify decomposable valuations as precisely the class for which W-Mon characterizes implementability, thus mirroring convex domains in IPV settings. 

\subsection{Beyond Decomposable Valuations}
\label{sub:beyond-decompose}

For general valuations beyond decomposable, a more complex condition for characterizing implementability than $f$-single-crossing is needed. The explanation for this is as follows:
In the case of decomposable valuations~$\vprofile$, the decomposability property naturally ensures that for every $i$,  $\sminus$ and outcomes $a,b\in \Delta(A)$, either $\pdif{v_i}{s_i}(a;s_i,\sminus) \le \pdif{v_i}{s_i}(b;s_i,\sminus)$ for all $s_i \in S_i$, or $\pdif{v_i}{s_i}(a;s_i,\sminus) \ge \pdif{v_i}{s_i}(b;s_i,\sminus)$ for all $s_i \in S_i$. In other words, 
the slopes $\pdif{v_i}{s_i}$ induce a global ordering of the outcomes that does not depend on $s_i$. This will not necessarily hold for general valuations and so $f$-single-crossing is ill-defined for such valuations.
We thus introduce the following definition and show it is the ``right'' generalization of $f$-single-crossing, in the sense that it is both necessary and sufficient for implementability. 

\begin{definition}\label{defn:weak-single-crossing}
 Let $f$ be a social choice function. We say that the valuations $\vprofile$ satisfy \emph{weak} $f$-single-crossing if for each agent $i$ and signal profile $\profile \in \mathcal S$, and for every $z \in S_i$ 
 it holds that:
 \begin{equation}\label{eq:weak-single-crossing-condition}
    \left\langle v_i(z,\sminus) - v_i(\profile), \; f(z,\sminus)  \right\rangle \geq \int_{s_i}^{z} \left\langle \pdif{v_i}{s_i} (t, \sminus), \; f(t, \sminus)  \right\rangle dt
 \end{equation}
\end{definition}


The above definition can be loosely viewed as follows. For $z>s_i$, the outcome $f(z,\sminus)$ must have a marginal improvement in $v_i$ --- as we increase the signal from $z$ to $s_i$ --- which is greater than the cumulative marginal improvements made by all outcomes $f(t,\sminus)$ for all $t\in[s_i,z]$. 

\begin{proposition}[Weak $f$-single-crossing characterizes implementability]
\label{prop:single-crossing-characterization}
    For any IDV social choice setting, a mechanism is ex-post IC-IR if and only if for every $i$, $\sminus$, weak $f$-single-crossing holds, and the following payment identity and payment inequality hold:
    \begin{align}
        p_i(\profile) &= \; p_i(0,\sminus) + \int_{0}^{s_i} \left\langle v_i(t,\sminus), \pdif{f}{s_i}(t,\sminus) \right\rangle dt ; \label{eq:payments1-single-crossing} \\
        p_i(0,\sminus) &\leq  \left\langle v_i(0,\sminus), f(0,\sminus) \right\rangle. \label{eq:payments2-single-crossing}
    \end{align}
    Moreover, if we additionally require prices to be non-negative, it is sufficient to additionally have the following payment identity:
    $
        p_i(0,\sminus) =  \left\langle v_i(0,\sminus), f(0,\sminus) \right\rangle.
    $
\end{proposition}
It is not hard to see that, for any decomposable valuation profile $\vprofile$, $f$-single-crossing implies weak $f$-single-crossing. 
In fact, under decomposable valuations, the two notions coincide, as we show in the following lemma.


\begin{lemma}\label{lem:sc-implies-weak-sc}
    A decomposable valuation profile $\vprofile$ satisfies $f$-single-crossing if and only if $\vprofile$ satisfies \emph{weak} $f$-single-crossing.
\end{lemma}
%
%


\subsection{Implementability in IDV versus IPV}\label{sec:idv-ipv-equiv}

In this section we show the parallels between single-crossing-based characterizations of implementability, and classic characterizations including C-Mon and W-Mon (see Appendix~\ref{sub:prelim-IC-IPV} for the classic definitions). Our main results in this section are Corollary~\ref{cor:weakSC-equiv-to-CMON} and Corollary~\ref{cor:SC-equiv-to-WMON},
and we achieve these using the Chung-Ely~\cite{CE02} framework (recall Lemma~\ref{lem:Chung-Ely} that transfers from IPV to IDV) as well as the results of~\citet{ABMH10}. See also Figure~\ref{fig:high-level-scheme} for an illustration of these connections.

We begin by defining generalizations of W-Mon and C-Mon for IDV settings.

\begin{definition}[W-Mon Generalized to IDV]
 Given any IDV social choice setting  with valuations $\vprofile$, a social choice function $f$ satisfies weak monotonicity (W-Mon) if for any $\profile\in \mathcal{S}$, player $i$ and signal $s'_i \in S_i$, the following holds: $f(\profile)=a$ and $f(s'_i,\sminus)=b$ implies that $v_i(b;s'_i,\sminus) - v_i(b;\profile) \ge v_i(a;s'_i,\sminus) - v_i(a;\profile)$.
\end{definition}

\begin{definition}[C-Mon Generalized to IDV]
     Given any IDV social choice setting with valuations $\vprofile$, a social choice function $f$ satisfies cycle monotonicity  (C-Mon) if for any player $i$, $\sminus$, and any set of $\ell+1$ signals $s^{(1)}_i, \ldots, s^{(\ell+1)}_i \in S_i$ with $s^{(\ell+1)}_i=s^{(1)}_i$, we have that
     $
        \sum_{k=1}^{\ell} \langle v_i(s^{(k)}_i,\sminus), f(s^{(k)}_i,\sminus) - f(s^{(k+1)}_i,\sminus) \rangle \geq 0.
     $ 
\end{definition}

The definition of a convex domain $\mathcal{V}$ 
in IPV settings translates to the following definition in IDV settings using the Chung-Ely framework.

\begin{definition} [Convex Domain for IDV]
Given any IDV social choice setting $(n,\mathcal{S},\mathcal{A},\vec v)$, for each $i$, $\sminus$, let $V_i(\sminus) = \{v_i(\profile)\in \reals^\mu \mid  s_i \in S_i\}$.
We say that the domain is convex if the closure of $V_i(\sminus)$ is convex for each $i$ and $\sminus$.
\end{definition}


With these definitions we get the following proposition from Theorem~\ref{thm:ipv-charaterization} and Lemma~\ref{lem:Chung-Ely} (\citet{ABMH10} and~\citet{CE02}).

\begin{proposition}
[Implementability with IDV]\label{prop:IDV-implementability}
Consider an IDV social choice setting $(n,\mathcal{S}, \mathcal{A},\vec v)$ and a social choice function $f$.
\begin{enumerate}
    \item(Cycle monotonicity is necessary and sufficient) $f$ is implementable if and only if it satisfies cycle monotonicity.
    \item(Weak monotonicity is necessary) Every implementable $f$ satisfies weak monotonicity.\footnote{For an alternative proof see Appendix~\ref{apx:proofs-for-sec-3}, Lemma~\ref{lem:wmon-necessary}.}
    \item(Weak monotonicity is sometimes sufficient) Suppose that the domain is convex, then every finitely-valued $f$ that satisfies weak monotonicity is implementable.
    \end{enumerate}
\end{proposition}

Putting together Proposition~\ref{prop:single-crossing-characterization} and bullet~(1) of Proposition~\ref{prop:IDV-implementability} we immediately get the following equivalence of weak-$f$-single-crossing and C-Mon.

\begin{corollary}
    \label{cor:weakSC-equiv-to-CMON}
For any IDV social choice setting with valuation profile $\vprofile$, it holds that a social choice function $f$ satisfies cycle monotonicity (C-Mon) if and only if $\vprofile$ satisfies weak $f$-single-crossing.
\end{corollary}

\vspace{1ex}
\noindent {\bf Decomposable valuations as convex domains. }
By definition, $V_i(\sminus)$ is a curve in $\reals^\mu$, and a curve can be a convex domain if only if it is a straight line. Therefore, it is easy to see that \linear valuations give rise to convex domains.
Less obviously, any decomposable (and continuous) valuation profile gives rise to a convex domain. The reason is that much like in \linear valuations, the \emph{direction} of the tangent to the curve $V_i(\sminus)$ remains constant for all signals $s_i\in S_i$ in decomposable valuations. 
We formalize the intuition of the discussion above in the following lemma.

\begin{lemma}
\label{lem:convex-decomposable}
Suppose $S_i$ is an interval in $\reals^+$ and $v_i$ is continuous in $s_i$. 
For each $i$, $\sminus$, let $D = \{v_i(\profile) \in \reals^\mu \mid s_i \in S_i \}$ be the induced domain, then (the closure of) $D$ is convex if and only if $v_i$ is decomposable.
\end{lemma}
\begin{proof}
    Fix some agent $i$ and signals $\sminus$.
    Suppose the domain $D$ is convex. We first establish the following claim; see proof in the appendix. 
    \begin{claim}\label{clm:convex-decomposable}
        For every two signals $s$ and $s'$ in $S_i$ and every $t\in [s,s']$ there exists some $\lambda \in [0,1]$ such that $v_i(t,\sminus) = (1-\lambda) \cdot v_i(s,\sminus) + \lambda \cdot v_i(s,\sminus)$. 
    \end{claim}
    
    We next show that for any signal $s \in S_i$ there exists $\lambda(s) \geq 0$ such that $v_i(s,\sminus) = (1-\lambda(s))v_i(0,\sminus) + \lambda(s) v_i(1,\sminus)$. For $s\in [0,1]$ it follows directly from the claim above. Consider $s>1$. By the claim above, there exists $\lambda'\in(0,1]$ such that 
    $
        v_i(1,\sminus) = (1 -\lambda') v_i(0,\sminus) + \lambda'v_i(s,\sminus)
    $. 
    Now by taking $\lambda(s)=1 - 1/\lambda'$ we obtain the desired equality.
    
    We conclude that $v_i$ can be written as $v_i(\profile) = \lambda(s_i)( v_i(1,\sminus)-v_i(0,\sminus)) + v_i(0,\sminus)$, which is a decomposable valuation with $\hat v_i(\profile) = \lambda(s_i)$, $h_i(\sminus) = v_i(1,\sminus)-v_i(0,\sminus)$ and $g_i = v_i(0,\sminus)$. This proves the forward direction of the lemma.
    
    For the converse direction, let $v_i(\profile) = \hat{v}_i(\profile)\cdot h_i(\sminus) + g_i(\sminus)$ and consider two signals $s,s'\in S_i$. For any $\lambda\in [0,1]$, we have
    $ 
     \lambda v_i(s,\sminus) + (1-\lambda)v_i(s,\sminus) = (\lambda \hat{v}_i(s,\sminus) + (1-\lambda)v_i(s',\sminus))\cdot h_i(\sminus) + g_i(\sminus).
    $
    
    Since $\hat{v}_i(\cdot,\sminus)$ is a continuous function from $S_i$ to $\reals$, by the intermediate value theorem there exists some $t\in S_i$ such that $\hat{v}_i(t,\sminus) = \lambda \hat{v}_i(s,\sminus) + (1-\lambda)\hat{v}_i(s',\sminus)$. Plugging this back into the decomposed formulation of $v_i$, we obtain 
    $
        v_i(t,\sminus) = \lambda \hat{v}_i(s,\sminus) + (1-\lambda)v_i(s',\sminus).
    $
    This proves that $D$ is a convex domain, concluding the proof of the lemma.
\end{proof}

\noindent {\bf Connecting $f$-single-crossing and W-Mon. }
The following lemma now follows from 
bullets (2) and (3) of Proposition~\ref{prop:IDV-implementability}, combined with Lemma~\ref{lem:convex-decomposable}.
 
\begin{lemma}\label{lem:wmon-implies-truthful-when-decomp}
Given any decomposable valuations $\vprofile$, a finite-valued social choice function $f$ is 
truthfully implementable if and only if $f$ satisfies weak monotonicity (W-Mon).
\end{lemma}

Recall that we {have already established} a stronger result in the previous section: in Theorem~\ref{thm:linear-characterization} we showed that for \emph{any} social choice function (potentially infinitely-valued), $f$-single-crossing is both necessary and sufficient for implementability under decomposable valuations. 

By Lemma~\ref{lem:wmon-implies-truthful-when-decomp} and Theorem~\ref{thm:linear-characterization}, 
we establish an equivalence between weak monotonicity and $f$-single-crossing under decomposable valuations for finite-valued $f$s.
The same holds for general $f$s by Lemmas~\ref{lem:sc-implies-weak-sc},~\ref{lem:weak-f-sc-to-wmon}, and \ref{lem:wmon-implies-f-sc-when-decomp}.

\begin{corollary}
\label{cor:SC-equiv-to-WMON}
For any IDV social choice setting with decomposable valuations $\vprofile$, it holds that a social choice function $f$ satisfies weak monotonicity (W-Mon) if and only if $\vprofile$ satisfies $f$-single-crossing  
\end{corollary}

\subsection{Decomposable Valuations form the Frontier of W-Mon Truthfulness}\label{sub:impossibility-for-non-decomposable}

An interesting implication that arises from the connection between decomposable valuations and convex domains is that decomposable valuations form the frontier of W-Mon truthfulness. The following theorem from \cite{ABMH10} shows that in IPV settings, when the domain is non-convex then W-Mon is insufficient for implementability.

\begin{theorem}[\citet{ABMH10}, Theorem 3]\label{thm:ashlagi-non-convex}
    Given any single-agent IPV social choice setting with a non-convex domain (which is not single-dimensional), there exists a finite-valued social choice function $f$ for which weak monotonicity (W-Mon) holds and yet $f$ is not implementable.
\end{theorem}

Lemma~\ref{lem:convex-decomposable} shows that for any valuation function $v$, the induced domain is convex if and only if $v$ is decomposable. Therefore, by applying Theorem~\ref{thm:ashlagi-non-convex} we obtain the following result. 


\begin{corollary}
\label{cor:impossibility-for-non-decomposable}
    For any non-decomposable valuation $v$ (which is not single-dimensional), there exists a social choice function $f$ for which weak monotonicity (W-Mon) holds and yet $f$ is not implementable.
\end{corollary}

This result implies that the class of decomposable valuations is the {\em frontier} of valuations in IDV settings for which weak monotonicity characterizes truthfulness. That is, for any valuation function that is non-decomposable, there exists a single agent example with some social choice function $f$, for which weak monotonicity is insufficient for implementability. Therefore, any class of valuations broader than decomposable valuations would require stronger conditions for implementability.

%% file: appendix.tex
\section{Preliminaries}
\subsection{Implementability with IPV}
\label{sub:prelim-IC-IPV}


The literature provides \emph{monotonicity} characterizations of implementability for both single-dimensional settings and general settings.
Monotonicity refers to how the social choice changes with changes in an agent's values.
In single-dimensional settings, it is well-known that $f$ is (dominant strategy IC-IR) implementable if and only if for every $i,\vprofile_{-i}$ it holds that $f_i(v_i,\vprofile_{-i})$ is monotone non-decreasing in $v_i$~\cite{Myerson1981}.
In multi-dimensional social choice settings, 
the following is a counterpart to single-dimensional monotonicity:


\begin{definition}[Weak monotonicity~\citep{LMN03}]
\label{def:WMON}
Consider a social choice setting $(n,\mathcal{V},\mathcal{A})$. A social choice function $f$ satisfies \emph{weak monotonicity} if for every $\vprofile\in \mathcal{V}$, agent $i\in[n]$ and $v'_i \in V_i$:
\[
    \left\langle v_i - v'_i , f(\vprofile) - f(v'_i,\vprofile_{-i}) \right\rangle \ge 0.
\]
\end{definition}
In other words, $f(\vprofile) = a$ and $f(v'_i,\vprofile_{-i}) = b$ for two alternatives $a,b\in\mathcal{A}$ implies that $v_i(a) - v'_i(a) \ge v_i(b) - v'_i(b)$ (equivalently, $v'_i(b) - v'_i(a) \ge v_i(b) - v_i(a)$). Intuitively, if switching from $v_i$ to $v'_i$ caused the social choice to switch from $a$ to $b$, then agent $i$'s value for $b$ relative to $a$ must have grown with the switch.

Weak monotonicity is a \emph{necessary} condition for a social choice function to be truthfully implementable (in any domain). 
However, unlike monotonicity in single-dimensional settings, weak monotonicity is \emph{insufficient} in many domains (and hence does not characterize truthfulness). A central line of work~\citep{LMN03,SY05,ABMH10} studies under which domains weak monotonicity is both sufficient and necessary for implementation. In particular, \citet{ABMH10} show that when (the closure of) the domain $\mathcal{V}$ is convex weak monotonicty is both sufficient and necessary for truthful implementation of any finite-valued social choice function $f$.

\citet{Rochet87} found that a different notion called \emph{cycle monotonicity} characterizes implementation for \emph{all} domains.
However this notion is less intuitive and considered much harder to work with.

\begin{definition}[Cycle monotonicity~\citep{Rochet87}]
Consider a social choice setting $(n,\mathcal{V},\mathcal{A})$. A social choice function~$f$ satisfies \emph{cycle monotonicity} if for every $i\in[n]$,\;  $\vprofile_{-i} \in \mathcal{V}_{-i},\; \ell\ge 2$ and ${v}^{(1)},{v}^{(2)},\ldots,{v}^{(\ell)} \in V_i$ where ${v}^{(\ell +1)} = {v}^{(1)}$, we have:
 \[
     \sum_{j=1}^{\ell} \langle {v}^{(j)} - {v}^{(j+1)}, f(v^{(j)},\vprofile_{- i}) \rangle \ge 0.
 \]
\end{definition}


\begin{observation}\label{obv:cmon-implies-wmon}
    Cycle monotonicity implies weak monotonicty, by setting $\ell = 2$.
\end{observation}

The following theorem summarizes the characterizations that are of particular interest to us.


\begin{theorem}[Implementability with IPV \citep{Rochet87,LMN03,SY05,ABMH10}]\label{thm:ipv-charaterization}
Consider a social choice setting $(n,\mathcal{V},\mathcal{A})$ and a social choice function $f$.
\begin{enumerate}
    \item (Cycle monotonicity is necessary and sufficient) $f$ is implementable if and only if it satisfies cycle monotonicity.
    \item (Weak monotonicity is necessary) Every implementable $f$ satisfies weak monotonicity.
    \item (Weak monotonicity is sometimes sufficient) Suppose $\mathcal{V}$ is a convex domain, then every finitely-valued $f$ that satisfies weak monotonicity is implementable.
    \end{enumerate}
\end{theorem}

\section{Proofs from Sections~\ref{sec:linear-characterization} and~\ref{sec:general-charac}}\label{apx:proofs-for-sec-3}

We first observe that Proposition~\ref{prop:single-crossing-characterization} is a strict generalization of Proposition~\ref{prop:linear-characterization} and Theorem~\ref{thm:linear-characterization}. In particular, under decomposable valuations, $f$-single crossing and weak $f$-single crossing are equivalent. Moreover, by Observation~\ref{obv:f-sc and strong-sc}, under decomposable valuations, strong single-crossing is equivalent to $f$-single crossing for a welfare maximizing social choice function $f$.  Therefore, it suffices to prove the more general result of Proposition~\ref{prop:single-crossing-characterization}.

\begin{proof}[Proof of Proposition~\ref{prop:single-crossing-characterization}. {[Following \citet{RT16}, Prop. 5.1]}]
    We first prove that for any ex-post IC and ex-post IR mechanism $(f,p)$, the payments $p$ must satisfy the payment identity and payment inequality. By definition of ex-post IC, for any agent $i$, profile $\profile\in \mathcal S$, and signal $t\in S_i$, the following inequalities must hold:
    \begin{align*}
        \langle v_i(\profile), f(\profile) \rangle - p_i (\profile) &\geq \langle v_i(\profile), f(t,\sminus) \rangle - p_i (t,\sminus) \\
        \langle v_i(t,\sminus), f(t,\sminus) \rangle - p_i (t,\sminus) &\geq \langle v_i(t,\sminus), f(\profile) \rangle - p_i (\profile)
    \end{align*}
    Rearranging and combining the two inequalities we obtain:
    \begin{align*}
        \langle v_i(\profile), f(t,\sminus) - f(\profile) \rangle  \leq p_i (t,\sminus) - p_i (\profile) \leq 
        \langle v_i(t,\sminus), f(t,\sminus) - f(\profile)\rangle
    \end{align*}
    Dividing by $t-s_i$ and taking the limit as $t$ goes to $s_i$ we obtain:
    \begin{align*}
        \pdif{p_i}{s_i} (\profile) =
        \langle v_i(\profile), \pdif{f}{s_i}(\profile)\rangle
    \end{align*}
    Integrating both sides, by the fundamental theorem of calculus, we obtain:
    \begin{align}\label{eq:payments-identity}
        p_i(\profile) = C +
        \int_{0}^{s_i}\langle v_i(t,\sminus), \pdif{f}{s_i}(t,\sminus)\rangle dt,
    \end{align}
    where $C$ is some arbitrary constant. Noting that plugging $s_i = 0$ into Equation \eqref{eq:payments-identity} yields $p_i(0,\sminus) = C$, we obtain the payment identity of Equation \eqref{eq:payments1-single-crossing}. This shows that condition  \eqref{eq:payments1-single-crossing} must hold for any ex-post IC mechanism.
    
    The payment inequality of Equation \eqref{eq:payments2-single-crossing} follows directly from the assumption that the mechanism $(f,p)$ is ex-post IR. Namely, by ex-post IR, the price agent $i$ pays given the signal profile $(0,\sminus)$ cannot be larger than agent $i$'s value for the signal profile $(0,\sminus)$. This shows that condition  \eqref{eq:payments2-single-crossing} must hold for any ex-post IR mechanism.
    
    Note that conditions \eqref{eq:payments1-single-crossing} and \eqref{eq:payments2-single-crossing} must hold for any ex-post IR and ex-post IC mechanism, regardless of whether the social choice function satisfies weak $f$-single-crossing or not. We next show that payments satisfying \eqref{eq:payments1-single-crossing} and \eqref{eq:payments2-single-crossing} guarantee the mechanism $(f,p)$ is ex-post IR and ex-post IC if and only if we have weak $f$-single-crossing.
    
    We first show that the mechanism is ex-post IC if and only if weak $f$-single-crossing holds. That is, we want to show that for every agent $i$, signal profile $\profile \in \mathcal S$, and signal $z\in S_i$ the following inequality holds:
    
    \begin{align}
        \langle v_i(\profile), f(\profile) \rangle - p_i (\profile) \geq \langle v_i(\profile), f(z,\sminus) \rangle - p_i (z,\sminus) \label{eq:ic-condition}
    \end{align}
    Plugging the price identity of Equation \eqref{eq:payments1-single-crossing} into Equation \eqref{eq:ic-condition} and rearranging we obtain:
    \begin{align*}
        \int_{s_i}^{z} \langle v_i(t,\sminus), \pdif{f}{s_i}(t,\sminus)\rangle dt \geq \langle v_i(\profile), f(z,\sminus) - f(\profile) \rangle
    \end{align*}
    Applying integration by parts,
    \begin{align*}
        \langle v_i(t,\sminus), \; f(t,\sminus) \rangle \Big\vert_{s_i}^{z} - \int_{s_i}^{z} \langle \pdif{v_i}{s_i}(t,\sminus), f(t,\sminus) \rangle dt \geq \langle v_i(\profile), f(z,\sminus) - f(\profile) \rangle
    \end{align*}
    Rearranging,
    \begin{align*}
        \langle v_i(z,\sminus) - v_i(\profile), \; f(z,\sminus) \rangle \geq \int_{s_i}^{z} \langle \pdif{v_i}{s_i}(t,\sminus), f(t,\sminus) \rangle dt,
    \end{align*}
    which is exactly Equation \eqref{eq:weak-single-crossing-condition}, the condition for weak $f$-single-crossing. We conclude that ex-post IC holds if and only if we have weak $f$-single-crossing.
    
    We next consider ex-post IR. We would like to show that for every agent $i$ and every signal profile $\profile \in \mathcal S$
    \begin{align} \label{eq:ir-condition}
        p_i(\profile) \leq \langle v_i (\profile), f(\profile) \rangle
    \end{align}
    Recall that by Equation \eqref{eq:payments1-single-crossing}, we have
    \begin{align*}
        p_i(\profile) = \; p_i(0,\sminus) + \int_{0}^{s_i} \left\langle v_i(t,\sminus), \pdif{f}{s_i}(t,\sminus) \right\rangle dt
    \end{align*}
    Applying integration by parts and rearranging we obtain
    \begin{align*}
        p_i(\profile) = \; \langle v_i (\profile), f(\profile) \rangle - \big(\left\langle v_i(0,\sminus), f(0,\sminus) \right\rangle - p_i(0,\sminus)\big) - \int_{0}^{s_i} \langle \pdif{v_i}{s_i}(t,\sminus), f(t,\sminus) \rangle dt
    \end{align*}
    Note now that the term in parenthesis is non-negative by Equation \eqref{eq:payments2-single-crossing} and the integration is non-negative as valuations are monotone non-decreasing in all signals (and thus $\pdif{v_i}{s_i}(t,\sminus)$ is a vector of non-negative reals). Therefore, the right hand side is st most $\langle v_i (\profile), f(\profile) \rangle$, so Equation \eqref{eq:ir-condition} holds and thus the mechanism satisfies ex-post IR. 
    
    It remains to prove the final part of the proposition. Namely, that the payment identity  $p_i(0,\sminus) =  \left\langle v_i(0,\sminus), f(0,\sminus) \right\rangle$ is sufficient for payments to be non-negative. By ex-post IC, for every agent $i$ and signal profile $\profile \in \mathcal S$ we have
    \begin{align*}
        \langle v_i(0,\sminus), f(0,\sminus) \rangle - p_i (0,\sminus) \geq \langle v_i(0,\sminus), f(\profile) \rangle - p_i (\profile)
    \end{align*}
    Rearranging we obtain
    \begin{align*}
        p_i (\profile) \geq  p_i (0,\sminus) + \langle v_i(0,\sminus), f(\profile) \rangle - \langle v_i(0,\sminus), f(0,\sminus) \rangle
    \end{align*}
    Hence, for the payment under profile $\profile$ to be non-negative, i.e., for $p_i(\profile)\geq 0$ to hold, it suffices for the right hand side of the inequality above to be non-negative. Thus, it is sufficient to have 
    \begin{align*}
        p_i (0,\sminus) \geq \langle v_i(0,\sminus), f(0,\sminus) \rangle - \langle v_i(0,\sminus), f(\profile) \rangle
    \end{align*}
    
    As valuations are non-negative for all signals, we have that $ \langle v_i(0,\sminus), f(\profile) \rangle \geq 0$, and therefore the following inequality is sufficient for non-negative prices: 
    \begin{align*}
        p_i (0,\sminus) \geq \langle v_i(0,\sminus), f(0,\sminus) \rangle
    \end{align*}
    Combining the above inequality with the payment inequality of Equation \eqref{eq:payments2-single-crossing}, we obtain the payment identity
    \begin{align*}
        p_i (0,\sminus) = \langle v_i(0,\sminus), f(0,\sminus) \rangle
    \end{align*}
    as a sufficient condition for prices to be non-negative. This proves the final part of the proposition, concluding the proof of the proposition.
\end{proof}

\begin{proof}[Proof of Lemma~\ref{lem:sc-implies-weak-sc}]
    We first show that, for any decomposable valuation profile, $f$-single-crossing implies weak $f$-single crossing.
    
    Fix some agent $i$ and profile $\profile \in \mathcal S$. 
    Since $f$-single-crossing holds, for all $t > t' \in S_i$ we have the following,
    \begin{equation}
        \left\langle \pdif{v_i}{s_i}(\profile), \; f(t,\sminus) \right\rangle \geq      \left\langle \pdif{v_i}{s_i}(\profile), \; f(t',\sminus) \right\rangle \label{eq:f-sc-condition}
    \end{equation}
     This is because for decomposable valuations the ordering of the slopes $\pdif{v_i}{s_i}$ doesn't depend on $s_i$.
    
    Let $z\in S_i$ be some signals for agent $i$. If $z = s_i$, then Equation \eqref{eq:weak-single-crossing-condition} vacuously holds. If $z>s_i$, since $f$-single-crossing holds,
    we obtain the following inequality from Equation \eqref{eq:f-sc-condition}:
    \begin{equation}\label{eq:pointwise-domination1}
        \int_{s_i}^{z}\left\langle \pdif{v_i}{s_i} (t,\sminus), \; f(z,\sminus)  \right\rangle dt \geq \int_{s_i}^{z}\left\langle \pdif{v_i}{s_i} (t,\sminus), \; f(t,\sminus)  \right\rangle dt
    \end{equation}
    By the fundamental theorem of calculus, this translates to
    \begin{equation}\label{eq:pointwise-domination2}
        \left\langle v_i(z,\sminus) - v_i(\profile), \; f(z,\sminus)  \right\rangle \geq \int_{s_i}^{z}\left\langle \pdif{v_i}{s_i} (t,\sminus), \; f(t,\sminus)  \right\rangle dt
    \end{equation}
    and therefore Equation \eqref{eq:weak-single-crossing-condition} holds for all $z>s_i$.
    
    We next consider the case where $z<s_i$.
    Similarly to above, we obtain the following inequality from Equation~\eqref{eq:f-sc-condition}: 
    \begin{equation}\label{eq:pointwise-domination3}
         \left\langle v_i(\profile) - v_i(z,\sminus), \; f(z,\sminus)  \right\rangle \leq \int_{z}^{s_i}\left\langle \pdif{v_i}{s_i} (t,\sminus), \; f(t,\sminus)  \right\rangle dt
    \end{equation}
    Flipping the direction of integration and multiplying both sides of the inequality by -1 we obtain Equation \eqref{eq:weak-single-crossing-condition}. 
    This proves that $f$-single crossing implies weak $f$-single crossing.
    
    The reverse direction follows from Lemmas~\ref{lem:weak-f-sc-to-wmon} and~\ref{eq:wmon-implies-f-sc-when-decomp1}, thus concluding the proof of the lemma.
\end{proof}
\begin{lemma}\label{lem:weak-f-sc-to-wmon}
    Weak $f$-single-crossing implies weak monotonicity.
\end{lemma}
\begin{proof}
    By weak $f$-single-crossing, for every agent $i$, $\profile$ and $z\in S_i$, we have that
    \begin{align}\label{eq:weak-f-sc-to-wmon1}
        \left\langle v_i(z,\sminus) - v_i(\profile), \; f(z,\sminus)  \right\rangle \geq \int_{s_i}^{z} \left\langle \pdif{v_i}{s_i} (t, \sminus), \; f(t, \sminus)  \right\rangle dt
    \end{align}
    and reversing the roles of $z$ and $s_i$ in Equation~\eqref{eq:weak-single-crossing-condition} we also have
    \begin{align}\label{eq:weak-f-sc-to-wmon2}
        \left\langle v_i(\profile) - v_i(z,\sminus), \; f(\profile)  \right\rangle \geq \int_{s_i}^{z} \left\langle \pdif{v_i}{s_i} (t, \sminus), \; f(t, \sminus)  \right\rangle dt
    \end{align}
    Combining Equations \eqref{eq:weak-f-sc-to-wmon1} and \eqref{eq:weak-f-sc-to-wmon2} we obtain
    \begin{align*}
        \left\langle v_i(z,\sminus) - v_i(\profile), \; f(z,\sminus)  \right\rangle \geq \left\langle v_i(z,\sminus) - v_i(\profile), \; f(\profile)  \right\rangle.
    \end{align*}
    This proves that weak monotonicity holds.
\end{proof}

\begin{lemma}\label{lem:wmon-implies-f-sc-when-decomp}
    Let $\vprofile$ be decomposable valuations and let $f$ be a social choice function. In this setting, weak monotonicity implies $f$-single-crossing.
\end{lemma}
\begin{proof}
    $v_i$ is decomposable and therefore there exist functions $\hat v_i : \mathcal{S} \to \reals^+$, $h_i : \mathcal{A}\times \mathcal{S}_{-i} \to \reals^+$, and $g_i: \mathcal{A}\times \mathcal{S}_{-i} \to \reals$ such that for every $a,\profile$, $v_i(a;\profile) = \hat{v}_i(\profile)\cdot h_i(a;\profile) + g_i(a;\profile)$.
    By weak monotonicity, for every agent $i$, $\profile$ and $z\in S_i$, we have that
    \begin{align*}
        \left\langle h_i(\sminus)(\hat v_i(z,\sminus)-\hat v_i(s_i,\sminus)), \; f(z,\sminus)  \right\rangle \geq \left\langle h_i(\sminus)(\hat v_i(z,\sminus)-\hat v_i(s_i,\sminus)), \; f(\profile)  \right\rangle.
    \end{align*}
    Noting that the term $(\hat v_i(z)-\hat v_i(s_i))$ is a scalar, we obtain
    \begin{align*}
        (\hat v_i(z,\sminus)-\hat v_i(s_i,\sminus))\left\langle h_i(\sminus), \; f(z,\sminus)  \right\rangle \geq (\hat v_i(z,\sminus)-\hat v_i(s_i,\sminus))\left\langle h_i(\sminus), \; f(\profile)  \right\rangle.
    \end{align*}
    Dividing both sides of the inequality by $(\hat v_i(z,\sminus)-\hat v_i(s_i,\sminus))$ and multiplying both sides by $\pdif{\hat v_i}{s_i}(t,\sminus)$ for arbitrary $t\in S_i$ we obtain
    \begin{align}\label{eq:wmon-implies-f-sc-when-decomp1}
        \left\langle h_i(\sminus)\pdif{\hat v_i}{s_i}(t,\sminus), \; f(z,\sminus)  \right\rangle \geq \left\langle h_i(\sminus)\pdif{\hat v_i}{s_i}(t,\sminus), \; f(\profile)  \right\rangle.
    \end{align}
    Notice that $\pdif{v_i}{s_i}(t,\sminus) = h_i(\sminus)\pdif{\hat v_i}{s_i}(t,\sminus)$, and therefore by plugging $t=s_i$ (resp. $t=z$) into Equation \eqref{eq:wmon-implies-f-sc-when-decomp1} we obtain Equation \eqref{eq:f-sc-condition}, which is equivalent to the definition of $f$-single-crossing. This proves that $f$-single-crossing holds.
\end{proof}

\begin{lemma}\label{lem:wmon-necessary}
Given any valuations $\vprofile$, a social choice function $f$ is ex-post truthfully implementable only if $f$ satisfies (generalized) W-Mon.
\end{lemma}

\begin{proof}
To see why weak monotonicity follows directly from ex-post incentive compatibility, recall that a social choice function $f$ is ex-post IC implementable if there exists a price function $p:\mathcal{S} \to \reals$ such that
\[
    \langle v_i(\profile), f(\profile)\rangle - p_i(\profile) \geq
    \langle v_i(\profile), f(s_i',\sminus)\rangle - p_i(s_i',\sminus) \qquad \forall i \in [n], \; \forall \profile \in \mathcal{S}, \; \forall s'_i \in S_i.
\]
Writing the same inequality while reversing the order of $s_i$ and $s'_i$ and summing with the inequality above, we obtain
\[
    \langle v_i(s_i',\sminus), f(s_i',\sminus) - f(\profile)\rangle \geq
    \langle v_i(\profile), f(s_i',\sminus) - f(\profile) \rangle \qquad \forall i \in [n], \; \forall \profile \in \mathcal{S}, \; \forall s'_i \in S_i,
\]
which implies $f$ is weakly monotone by definition.
\end{proof}

\begin{proof}[Proof of Claim~\ref{clm:convex-decomposable}]
    Consider some outcome $a\in \mathcal A$. Notice that there exists $\lambda_a\in[0,1]$ such that $v_i(a; t,\sminus) = (1-\lambda_a)\cdot v_i(a; s,\sminus) + \lambda_a \cdot v_i(a; s',\sminus)$. Namely, 
    \[
        \lambda_a = \frac{v_i(a; s,\sminus) - v_i(a; t,\sminus)}{v_i(a; s,\sminus) - v_i(a; s',\sminus)}
    \]
    
    Assume towards contradiction that there exist two outcomes $a,b\in\mathcal A$ such that $\lambda_a\neq\lambda_b$. Wlog assume $\lambda_a < \lambda_b$. By convexity of $D$ and monotonicity of $v_i$, there exist $x_a<x_b\in [s,s']$ such that
    \begin{align*}
        v_i(x_a,\sminus) &= (1-\lambda_a) \cdot v_i(s,\sminus) + \lambda_a \cdot v_i(s',\sminus) \\
        v_i(x_b,\sminus) &= (1-\lambda_b) \cdot v_i(s,\sminus) + \lambda_b \cdot v_i(s',\sminus)
    \end{align*}
    By convexity there exists a signal $y\in (x_a,x_b)$, such that, $v_i(y,\sminus) = (v_i(x_a,\sminus)+v_i(x_b,\sminus))/2$. Now notice that $v_i(a;y,\sminus) > v_i(a;t,\sminus)$, and $v_i(b;y,\sminus) < v_i(b;t,\sminus)$. This is a  contradiction since $v_i$ is monotone in $s_i$. Hence $\lambda_a = \lambda_b$ for all $a,b\in \mathcal{A}$, proving the claim.
\end{proof}

\section{Local Exclusion under Multi-Parameter Signals}\label{app:local-exclusion}

In this section we show an example when local exclusion does not help with truthfulness even in the more general setting where each agent has multi-signals.
Under the multi-parameter signals setting, each agent $i$ has different signals $s_{i}^{(j)}$ for each project $j$, and the value of any agent $i$ for project $j$, $v_{i}{(j;\cdot)}$, only depends on the signals $\profile^{(j)}$, where $\profile^{(j)} = (s_1^{(j)}, s_2^{(j)},\ldots, s_n^{(j)})$.\footnote{Note that, the focus of this paper is the single-parameter signal setting where $s_{i}^{(j)} = s_i$ for all $i,j$.}

Recall that under global exclusion, when agent $i$ is excluded (that is, obtains no allocation from the mechanism), there is no incentive for $i$ to misreport $s_i$. However, even with multi-parameter signals, under local exclusion (i.e., each project $j$ may be associated with different excluded agents $E_j$) an agent $i\in E_j$ may have an incentive to misreport $s_i^{(j)}$.

Consider a setting with two agents $i\in \{1,2\}$, three projects $j\in \{1,2\}$, and $k =1$. Only agent $1$ has signals $s_1^{(1)},s_1^{(2)} \in \{0,1\}$. The valuations of the agents are,

\begin{align*}
    v_1(1;s_1^{(1)}) =&\; \varepsilon  &v_1(2;s_1^{(2)}) =&\; H\cdot s_1^{(2)} + 1\\
    v_2(1;s_1^{(1)}) =&\; H\cdot s_1^{(1)} &v_2(2;s_1^{(2)}) =&\; 0
\end{align*}


Suppose $E_1 = \{1\}$ and $E_2 =\{2\}$, that is, agent $1$ is excluded from project $1$ and agent $2$ is excluded from project $2$. If $s_1^{(1)} =0$ to achieve any approximation to the social welfare a deterministic mechanism will always allocate project $2$ (to agent $1$), hence agent $1$ will misreport $s_1^{(1)}=0$.